\documentclass[12pt,preprint]{aastex}

\shorttitle{Spitzer/JCMT Observations of NGC 4594}
\shortauthors{Bendo et al.}

\begin{document}

\title{{\it Spitzer} and {\it JCMT} Observations of the Active
Galactic Nucleus in the Sombrero Galaxy (NGC~4594)}

\author{George J. Bendo\altaffilmark{1, 2}, 
Brent A. Buckalew\altaffilmark{3}
Daniel A. Dale\altaffilmark{4},
Bruce T. Draine\altaffilmark{5}, 
Robert D. Joseph\altaffilmark{6},
Robert C. Kennicutt, Jr.\altaffilmark{7, 2},
Kartik Sheth\altaffilmark{3},
John-David T. Smith\altaffilmark{2},
Fabian Walter\altaffilmark{8},
Daniela Calzetti\altaffilmark{9},
John M. Cannon\altaffilmark{8},
Charles W. Engelbracht\altaffilmark{2},
Karl D. Gordon\altaffilmark{2},
George Helou\altaffilmark{3},
David Hollenbach\altaffilmark{10},
Eric J. Murphy\altaffilmark{11},
and H\'el\`ene Roussel\altaffilmark{8}}
\altaffiltext{1}{Astrophysics Group, Imperial College, Blackett Laboratory,
    Prince Consort Road, London SW7 2AZ, United Kingdom; 
    g.bendo@imperial.ac.uk}
\altaffiltext{2}{Steward Observatory, University of Arizona, 933 North 
    Cherry Avenue, Tucson, AZ 85721 USA.}
\altaffiltext{3}{California Institute of Technology, MC 314-6, Pasadena, CA 
    91101 USA.}
\altaffiltext{4}{Department of Physics and Astronomy, University of Wyoming, 
    Laramie, WY 82071 USA.}
\altaffiltext{5}{Princeton University Observatory, Peyton Hall, Princeton, 
    NJ 08544-1001 USA.}
\altaffiltext{6}{Institute for Astronomy, University of Hawaii, 
    2680 Woodlawn Drive, Honolulu, HI 96822 USA.}
\altaffiltext{7}{Institute of Astronomy, University of Cambridge, 
    Cambridge CB3 0HA, United Kingdom}
\altaffiltext{8}{Max-Planck-Institut f\"ur Astronomie, K\"onigstuhl 17, 
    D-69117 Heidelberg, Germany}
\altaffiltext{9}{Space Telescope Science Institute, 3700 San Martin Drive, 
    Baltimore, MD 21218 USA.}
\altaffiltext{10}{NASA Ames Research Center, MS 245-3, Moffett Field, CA 
    94035-1000 USA.}
\altaffiltext{11}{Department of Astronomy, Yale University, P.O. Box 208101, 
    New Haven, CT 06520 USA.}

\begin{abstract}
We present {\it Spitzer} 3.6-160 $\mu$m images, {\it Spitzer} mid-infrared
spectra, and {\it JCMT} SCUBA 850 $\mu$m images of the Sombrero Galaxy (NGC
4594), an Sa galaxy with a $10^9$ M$_\odot$ low luminosity active
galactic nucleus (AGN).  The brightest infrared sources in the galaxy
are the nucleus and the dust ring. The spectral energy distribution of
the AGN demonstrates that, while the environment around the AGN is a
prominent source of mid-infrared emission, it is a relatively weak
source of far-infrared emission, as had been inferred for AGN in
previous research.  The weak nuclear 160~$\mu$m emission and the
negligible polycyclic aromatic hydrocarbon emission from the nucleus
also implies that the nucleus is a site of only weak star formation
activity and the nucleus contains relatively little cool
interstellar gas needed to fuel such activity.  We propose that this
galaxy may be representative of a subset of low ionization nuclear
emission region galaxies that are in a quiescent AGN phase because of
the lack of gas needed to fuel circumnuclear star formation and
Seyfert-like AGN activity.  Surprisingly, the AGN is the predominant
source of 850~$\mu$m emission.  We examine the possible emission
mechanisms that could give rise to the 850~$\mu$m emission and find
that neither thermal dust emission, CO line emission, bremsstrahlung
emission, nor the synchrotron emission observed at radio
wavelengths can adequately explain the measured 850~$\mu$m flux
density by themselves.  The remaining possibilities for the source of
the 850~$\mu$m emission include a combination of known emission
mechanisms, synchrotron emission that is self-absorbed at wavelengths
longer than 850~$\mu$m, or unidentified spectral lines in the
850~$\mu$m band.
\end{abstract}


\section{Introduction}

A key yet simple point in understanding the infrared spectral energy
distributions (SEDs) of galaxies is separating the component of the
dust emission that is primarily heated by an active galactic nucleus
(AGN) from the component that is primarily heated by star formation.
Such simple knowledge is vital for examining the connection between
starbursts and AGN, particularly those that are dust-enshrouded; for
modeling the SEDs of dust emission from galaxies; for validating or
modifying the unified model of AGN; and for understanding distant,
unresolved AGN.  This issue is especially important in debates on
the origins of far-infrared emission in powerful AGNs such as
quasi-stellar objects \citep[see][for a discussion]{gc00}.

Results from the Infrared Astronomical Satellite (IRAS) have shown that
the 60~$\mu$m~/~25~$\mu$m flux ratios of Seyfert galaxies and other AGN
were lower than for starbursts and other spiral galaxies
\citep[e.g.][]{dmld85, od85, dml87, kl87, rrj87, rc89}.  The
enhancement in the 25~$\mu$m emission was interpreted as originating
from the environment of the AGN, whereas the 60 and 100~$\mu$m
emission was interpreted as originating from star formation activity.
However, since both the AGN and star formation were co-spatial,
disentangling the infrared SEDs of both has proved to be difficult.

A few studies that either combined ground-based mid-infrared and IRAS
data \citep[e.g.][]{metal95} or used Infrared Space Observatory data
\citep[e.g.][]{rp97,prs98,prf00} began to separate the infrared
emission from AGN and star formation in nearby galaxies. Most of the
evidence suggesting that the far-infrared emission is related to star
formation rather than AGN activity relies on the similarities between
the far-infrared SEDs of Seyfert and non-Seyfert galaxies, on the
consistency of the far-infrared SEDs in a broad range of Seyfert
galaxies, and on the relation between far-infrared emission and
stellar emission.  While these results are good evidence for the far-infrared
emission of Seyfert galaxies originating from circumnuclear star
formation, the results are only inferences.

Generally, it is very challenging to separate the infrared emission
from dust heated by an AGN and the infrared emission from dust heated
by circumnuclear star formation.  The problem lies not only with the
limited resolution of previous instrumentation.  The major problem is
that many nearby Seyfert galaxies have circumnuclear star formation
that will produce centralized dust emission that is difficult to
disentangle from far-infrared emission from dust heated by the AGN.
Both sources of emission are effectively superimposed.

The \object[NGC 4594]{Sombrero Galaxy (NGC 4954)} at a distance of
9.2~Mpc (the average of measurements from \citet{fetal96} and
\citet{aetal97}) is ideal for studying the separate SEDs of dust
heated by the AGN and dust heated by starlight.  The galaxy's nucleus
is classified in \citet{hfs97} as a low ionization nuclear emission
region (LINER).  It does contain a supermassive black hole with a mass
of $10^9$ M$_\odot$ \citep{kbaetal96}, and the AGN is detected as a
point source in hard X-ray \citep[e.g.][]{pffta02, pbfk03} and radio
\citep[e.g.][]{hvd84} emission.  What makes this particular LINER
unique is that the the geometry of the mid- and far-infrared dust
emission is relatively easy to model, and the dust that is primarily
heated by star formation is mostly concentrated in a ring relatively
far from the AGN, as shall be demonstrated in this paper.  Therefore,
the SED of the dust heated by the AGN can be separated from the SED of
the diffuse interstellar dust.  This galaxy has been studied in the
infrared/submillimeter waveband in previous works
\citep[e.g.][]{rlsnkldh88, skcs97, kwd05}, but these studies were
limited by the resolution of the IRAS data and could only examine the
global SED in the far-infrared.

In this paper, we present {\it Spitzer Space Telescope}
\citep{wetal04} Infrared Array Camera \citep[IRAC;][]{fetal04}
3.6-8.0~$\mu$m images, Multiband Imaging Photometer for Spitzer
\citep[MIPS;][]{ryeetal04} 24-160~$\mu$m images, and Infrared
Spectrograph \citep[IRS;][]{hetal04} mid-infrared spectra as well as
{\it James Clerk Maxwell Telescope} ({\it JCMT}) Submillimeter
Common-User Bolometer Array \citep[SCUBA;][]{hetal99} 850~$\mu$m
images of the Sombrero Galaxy that we will use to examine the separate
SEDs of the AGN and the dust ring.  In Section~\ref{s_obs}, we
describe the observations and data reductions.  In
Section~\ref{s_image}, we discuss the images qualitatively.  In
Section~\ref{s_spectra}, we briefly discuss the IRS mid-infrared
spectra.  In Section~\ref{s_sed}, we model the images of the galaxy in
each waveband to determine SEDs for global emission and for the separate
physical components of the galaxy.  Then, in
Section~\ref{s_discussion}, we discuss the major results from the SED
of the AGN.

\section{Observations and Data Reduction \label{s_obs}}

\subsection{3.6-8.0~$\mu$m Images}

The 3.6 - 8.0~$\mu$m data were taken with IRAC on the {\it Spitzer Space
Telescope} on 2004 June 10 in IRAC campaign 9 and on 2005 January 22 in
IRAC Campaign 18 as part of the SINGS survey \citep{ketal03}.  The
observations consist of a series of $5^\prime \times 5^\prime$
individual frames that are offset $2.^\prime5$ from each other.  The
two separate sets of observations allow asteroids to be recognized and
provide observations at orientations up to a few degrees apart to ease
removal of detector artifacts.  Points in the center are imaged eight
times in 30 s exposures.  The full-width half-maxima (FWHM) of the
point spread functions (PSFs) are
$1.^{\prime\prime}6-1.^{\prime\prime}9$ for the four wavebands.

The data are processed using a special SINGS IRAC pipeline.  First, a
geometric distortion correction is applied to the individual frames.
Data from the second set of observations are rotated to the same
orientation as the first set of observations.  Bias levels are
subtracted from the 5.7~$\mu$m data by subtracting a bias frame (made
by combining all data frames for the observations) from each frame.
Next, the image offsets are determined through image
cross-correlation.  Following this, bias drift is removed.  Finally,
cosmic ray masks are created using standard drizzle methods, and final
image mosaics are created using a drizzle technique.  The final pixel
scales are set at $\sim0.^{\prime\prime}75$.  A final background is
measured in small regions outside the target that are free of bright
foreground/background sources, and then this final background is
subtracted from the data.  The contribution of uncertainties in the
background (both in terms of the statistical fluctuations of the
pixels and the uncertainty in the mean background subtracted from the
data) to uncertainties in the integrated global flux densities is less
than 0.1\%.  The dominant source of uncertainty is the 30\%
uncertainty in the calibration factor (including the uncertainty in
the extended source calibration) applied to the final mosaics.

\subsection{24-160~$\mu$m Images}

The 24, 70, and 160~$\mu$m data were taken with MIPS on the {\it Spitzer
Space Telescope} on 2004 July 10 and 12 in MIPS campaign 10 as part of
the SINGS survey.  The observations were obtained
using the scan-mapping mode in two separate visits to the galaxy.  Two
separate sets of observations separated by more than 24~h allow
asteroids to be recognized and provide observations at orientations up
to a few degrees apart to ease removal of detector artifacts. As a
result of redundancy inherent in the scan-mapping mode, each pixel in
the core map area was effectively observed 40, 20, and 4 times at 24,
70, and 160~$\mu$m, respectively, resulting in integration times per
pixel of 160 s, 80 s, and 16 s, respectively.  The FWHM of the PSFs of
the 24, 70, and 160~$\mu$m data are $6^{\prime\prime}$,
$15^{\prime\prime}$, and $40^{\prime\prime}$, respectively.

The MIPS data were processed using the MIPS Data Analysis Tools
\citep{getal05} version 2.80.  The processing for the 24~$\mu$m data
differed notably from the 70 and 160~$\mu$m data, so they will be
discussed separately.

First, the 24~$\mu$m images were processed through a droop correction
(that removes an excess signal in each pixel that is proportional to
the signal in the entire array) and a non-linearity correction.
Following this, the dark current was subtracted.  Next,
scan-mirror-position dependent flats and scan-mirror-position
independent flats were applied to the data.  Latent images from bright
sources, erroneously high or low pixel values, and unusually noisy
frames were also masked out.  Finally, mosaics of the data from each
set of observations were made.  In each mosaic, the background was
subtracted in two steps.  First, to remove the broad zodiacal light
emission, a function that varied linearly in the x and y directions
was fit to the region outside the optical disk in a box three times
the size of the optical major axis of the galaxy.  This plane was then
subtracted from the data.  Next, to remove residual background
emission from cirrus structure near the galaxy, an additional offset
measured in multiple small circular regions near the optical disk was
subtracted.  After this final subtraction, the two mosaics were
averaged together to produce the final 24~$\mu$m mosaic.

In the 70 and 160~$\mu$m data, readout jumps and cosmic ray hits were
first removed from the data.  Next, the stim flash frames taken by the
instrument were used as responsivity corrections.  The dark current
was subtracted from the data, an illumination correction was applied,
and then short term variations in signal (i.e. short-term drift) were
subtracted from the data.  (This last step also subtracts off the
background.)  Following this, erroneously high or low pixel values
were identified statistically or visually and removed from the data.
Single 70 and 160~$\mu$m mosaics were made from all of the data, and a
residual offset measured in two regions to the north and south of the
target was subtracted from the final maps.

The background noise is a relatively small contributor (less than
0.1\%) to the uncertainties in the integrated global flux densities.  The
dominant source of uncertainty is the uncertainty in the calibration factors
applied to the final mosaic, which is 10\% at 24~$\mu$m and 20\% at 70
and 160~$\mu$m.

\subsection{850~$\mu$m Image}

The 850~$\mu$m data were taken with SCUBA at the {\it JCMT} on UT dates
2004 January 17-18.  Six maps (each of which were ten integrations, or
10.7 min of on-target integration) were taken in jiggle map mode, with
each map covering $2.^\prime3$ hexagonal regions.  Each map was
slightly offset from the others to ensure that no holes in the maps
were created by noisy bolometers.  The total map effectively covers
the inner $3^\prime \times 2.^\prime25$ region of the galaxy, with 60
integrations (64 min of on-target integration) of observing time of
the galaxy nucleus.  The FWHM of the PSF of the 850~$\mu$m data is
$15^{\prime\prime}$.

The 850~$\mu$m data were processed with the SCUBA User Reduction
Facility \citep{jl98}.  The data were first flatfielded and corrected
for atmospheric extinction.  Noisy bolometers were removed next,
followed by spike removal.  Then the background signal was subtracted
using the signal from several bolometers at the north and south edges
of the images.  The data were calibrated using observations of the
submillimeter standards CRL~618 and IRC~+10216 and then regridded onto
the sky plane.  An additional background subtraction was then
performed to remove any residual offset that appears when the
bolometer signals are regridded onto the sky plane.  The background
noise is a relatively small contributor to the uncertainties in the
integrated fluxes (approximately 1\%) .  The dominant source of uncertainty
is the 10\% uncertainty in the calibration factor applied to
the final image.

\subsection{Mid-Infrared Spectra}

The 5-38~$\mu$m spectra were taken with IRS on the {\it Spitzer Space
Telescope} on 2004 June 24 and 28 in IRS campaign 9 as part of the
SINGS survey.  The observations were taken in the spectral mapping
mode, in which the slit is moved in a raster pattern to build up a
redundantly-sampled spectral map of the target region.  The
observations presented here were made with the short-low (5-15~$\mu$m,
R=50-100), long-low (14-38~$\mu$m, R=50-100), and long-high
(20-37~$\mu$m, R=600) IRS modules.  The observations were made as a
series of pointings where the spectrometer's slits were stepped across
the center of the galaxy.  As a result, the observations produce
spectral cubes.  The sizes of the observed regions depend primarily on
the specific module used for the observation.

S12 processed BCD data were used for this analysis.  The data were
assembled into three-dimensional spectral cubes using Cubism
\citep[see][Section 6.2]{ketal03}.  When the spectral cubes are built,
bad pixels are identified and masked out, background subtraction is
performed, and flux calibration is applied to the data.  The calibration
uncertainties are 25~\%.

\section{Results}

\subsection{Images of NGC~4594 \label{s_image}}

Figure~\ref{f_images} shows the 3.6, 8, 24, 70, 160, and 850~$\mu$m
images of NGC~4594.  Throughout these wavebands, the sources of
emission consist mainly of a nuclear source, a ring with an
approximate semimajor axis of $\sim140^{\prime\prime}$
($\sim6.2$~kpc), a thin disk inside the ring composed of stars and
dust (referred to as the inner disk in this paper), and a bulge.  The
relative contributions of each of these components to each waveband
vary substantially.  Aside from these four components, no other
significant sources of emission are visible in the galaxy.

In nearby galaxies, the 3.6~$\mu$m band is dominated by stellar
emission \citep[e.g.][]{letal03}.  In NGC~4594, the bulge and the
inner disk are the two extended features with the highest surface
brightnesses at this waveband.  The ring is nearly invisible.

The 8~$\mu$m emission from spiral galaxies contains a mixture of
polycyclic aromatic hydrocarbon (PAH) emission, hot thermal emission
from very small grains, and the Rayleigh-Jeans tail of stellar
blackbody emission \citep[e.g.][]{hetal00, letal03}.  In the
observations of NGC~4594 in this waveband, the ring is a much more
prominent source of emission, which implies that it is a strong source
of PAH and hot dust emission. Note that the ring (at this resolution)
is relatively featureless; only a few knots of enhanced 8~$\mu$m
emission are visible in the ring.  Small extensions from the ring that
resemble spiral arms are just barely visible on the east and west ends
of the ring.  Emission from the inner disk is still visible in this
waveband.  The bulge emission, which originates from stars (and
possibly from hot dust in the atmospheres of evolved stars), is much
weaker than at 3.6~$\mu$m.

The 24~$\mu$m band mostly traces very small grain emission
\citep[e.g.][]{ld01}, although it could also include stellar emission
or thermal emission from strongly-heated large dust grains.  In this
waveband, the nucleus is a more prominent source of emission than at
shorter wavelengths.  The ring and the inner disk are also still
strong sources of emission.  Except for a bright knot in the southwest
portion of the ring, the ring contains no notable structures.  The
bulge almost disappears completely at this wavelength, although faint
traces of the emission are still visible to the north and south of the
ring.

The 70~$\mu$m band is one of two bands used in this paper that
primarily traces large grain emission \citep[e.g.][]{ld01}.  The
appearance of the 70~$\mu$m image looks similar to the 24~$\mu$m
image.  The nucleus, the inner dust disk, and the ring are all still
visible, but the bulge is indistinguishable from the background noise
(as is expected, since little cool interstellar dust is associated
with the bulges of galaxies).

In the 160~$\mu$m band, which also traces large grain emission
\citep[e.g.][]{ld01}, the ring is the most prominent source.  The
inner disk is a weak source of emission in this waveband, and it
cannot be easily distinguished from the ring.  In stark contrast to
the 24 and 70~$\mu$m images, the nucleus is almost completely
invisible.  These results suggest that the environment of the active
nucleus contains relatively little dust and that the dust that is
present is heated very strongly.

Even more surprising than the relatively weak 160~$\mu$m emission from
the nucleus is the relatively strong emission from the nucleus
observed at 850~$\mu$m.  Typically, the 850~$\mu$m emission traces
$\sim$20-30~K dust emission related to the 160~$\mu$m emission
\citep[e.g.][]{de01, betal03, rtbetal04}.  If these previous results
applied to this galaxy, then the ring would be the strongest feature
observed at 850~$\mu$m, and the nucleus would be relatively faint.
Instead, the nucleus is the only submillimeter source detected above
the $3\sigma$ level, which implies that the submillimeter emission
arises from some source other than $\sim$20-30~K dust.

To understand these images better, we will need to create models of
the images that include components for each of the images.  The
results of fitting the models to the data will include flux densities
that can be used to construct SEDs.  This is done in
Section~\ref{s_sed}.  First, however, we will look at the results from
the mid-infrared spectroscopy.

\subsection{Mid-Infrared Spectra of the Nucleus and Ring \label{s_spectra}}

We can further examine the nature of the nucleus and ring of this
galaxy using the IRS spectra.  The goal is to understand the primary
sources of the dust heating in the nucleus and ring, and specifically
to understand if the dust in the nucleus is heated by star formation
or by the AGN.  An AGN is clearly present in this galaxy, since a
central synchrotron source has been detected in radio
\citep[e.g.][]{hvd84} and X-ray \citep[e.g.][]{pffta02, pbfk03}
observations and a $10^9$~M$_\odot$ central object has been
detected through stellar dynamics \citep{kbaetal96}.  The results for
this galaxy will reveal whether the AGN is the dominant energy source
for the nucleus or if star formation is a significant factor.  These
data can also be used as a test of mid-infrared spectral line
diagnostics on the spectrum of a known low-luminosity AGN.

The regions in which spectra were extracted are shown in
Figure~\ref{f_extregions}.  Spectra for individual regions in the
galaxy are shown in Figure~\ref{f_spectra} and \ref{f_spectra_o4}.
Notable features measured in the spectrum of the AGN are listed in
Table~\ref{t_spectralmeas}.

The nuclear spectrum is consistent with what is expected from an AGN.
Metal lines such as [Ne~{\small II}], [Ne~{\small III}], [O~{\small
IV}], and [S~{\small III}] are present, but PAH features (such as the
features at 6.2, 7.7, 11.3, 16.4, and 17.1~$\mu$m are weak or absent.
Interestingly, some high excitation lines, notably the [Ne~{\small V}]
line at 14.3~$\mu$m, are also absent.  This is not surprising, as this
is a LINER nucleus with low levels of ionization, although [Ne~{\small
V}] emission often corresponds to AGN activity.  The absence of
these high excitation lines in this galaxy's spectrum may have
implications for using high excitation lines as a diagnostic for
detecting AGNs in other LINERs.  Aside from the absence of high
excitation lines, the spectral lines are consistent with AGN activity.
In the diagnostic diagrams of \citet{getal98} and \citet{pst04}, the
high [O~{\small IV}]~/~[Ne~{\small II}] ratio and the weak 6.2 and
7.7~$\mu$m PAH features indicate that this is a source dominated by
AGN activity.  In the alternate diagnostic diagram of D. A. Dale et
al. (2006, in preparation), where the 34.8~$\mu$m [Si~{\small II}] is
used in place of the 25.9~$\mu$m [O~{\small IV}] line, the nucleus
also falls in the AGN dominated regime.

The spectrum of the ring is more consistent with star formation
activity than the nucleus.  PAH emission from the ring, especially in
the 11.3~$\mu$m band, is much stronger.  Qualitatively, the stronger
PAH features would shift this galaxy into the star formation regime in
the diagnostic diagrams of \citet{getal98}, \citet{pst04}, and
D. A. Dale et al. (2006, in preparation).  Unfortunately, the
observations are too limited to show that the spectrum of the ring is
consistent with star formation.  The long-high observations, which are
needed to measure the 25.9~$\mu$m [O~{\small IV}] line, do not
adequately cover the ring.  The signal-to-noise ratio is too low for
the low-resolution spectrum between 5 and 8~$\mu$m for accurate
measurement of the 6.2 and 7.7~$\mu$m PAH features.  Finally, because
of the limitations of the spatial resolution, a significant fraction
of the flux at wavelengths greater than 20~$\mu$m may include emission
from the AGN, thus making measurements of the 25.9~$\mu$m [O~{\small
IV}] and 34.8~$\mu$m [Si~{\small II}] lines suspect.

Images of the 11.3~$\mu$m PAH feature and 25.9~$\mu$m [O~{\small IV}]
line emission in Figure~\ref{f_imgline} bolster these results.  These
images were made from the IRS data.  To produce the images, the
continuum was identified and subtracted on a pixel-by-pixel basis,
leaving just the spectral feature emission.  In the resulting map for
the PAH feature, the emission originates entirely from the ring of the
galaxy.  Even if PAH emission is treated as only a qualitative tracer
of star formation, these results suggest that the star formation is
confined to the ring.  The [O~{\small IV}] emission, however,
originates mostly from the nucleus.  These results from the PAH and
[O~{\small IV}] maps are consistent with the interpretation of an AGN
dominating the energetics of the nucleus.

Although generally the absence of nuclear PAH emission implies weak
star formation activity \citep[e.g.][]{getal98, pst04}, PAH emission
may be absent in some situations where strong star formation is
present.  PAH emission is absent in low metallicity star formation
regions \citep[e.g.][]{egrwdl05}.  This could be an explanation for
the absence of PAH emission from the nucleus of NGC~4594.  However,
since the metallicity of the ISM usually peaks in the centers of
galaxies and since PAH emission is seen in the ring, the metallicity
of the center is probably above the metallicity threshold necessary
for PAH emission.  PAH emission may also be absent in star formation
regions because of the presence of hard radiation fields
\citep[e.g.][]{metal05}.  If this is the case for NGC~4594, the hard
radiation field would be from the AGN, which is the dominant source of
X-ray photons in the nuclear environment \citep{pffta02, pbfk03}.  So,
even if circumnuclear star formation is present, the AGN still
dominates the interstellar radiation field in the vicinity of the
nucleus. Placed in the context of dust heating, the weak nuclear PAH
emission indicates that the dust emission is primarily heated by the
AGN.

\subsection{Broadband Spectral Energy Distributions for the Different 
Spatial Components \label{s_sed}}

To further investigate the dust heated by the AGN (which we will refer
to as the AGN emission for simplicity), we will examine the 3.6 -
850~$\mu$m broadband SEDs.  First, it is necessary to separate the
emission from the AGN from the various other spatial components in the
galaxy, so we will first model the 3.6-160~$\mu$m images in
Section~\ref{s_imagemodel}.  The parameters from the models fit to the
data include flux densities that are then used to construct SEDs.
These SEDs are analyzed in Section~\ref{s_sedanalysis}.

\subsubsection{Image Modeling \label{s_imagemodel}}

The basic image models consists of four parts: an unresolved nuclear
point source (from the region heated by the AGN), a dust ring with a
radius and radial width described by the Gaussian function
$e^{-\frac{(r-r_0)^2}{2w^2}}$, an inner dust disk with an exponential
profile described as $e^{-\frac{r}{h}}$ that was truncated at the
radius of the ring, and a bulge with a simplified de Vaucouleurs
profile $e^{-7.67 r_{ell}^{0.25}}$, where $r_{ell} = (
(\frac{x}{r_x})^2 + (\frac{y}{r_y})^2 )^{0.5}$ (with $r_x$ and $r_y$
representing the semimajor and semiminor axes of the ellipse that
contains half of the total light from the bulge).  Each component also
has a scaling term that gives the total flux density of the component.
These components were convolved to the resolutions of the various {\it
Spitzer} wavebands with PSFs created with STinyTim\footnote{Available
from
http://ssc.spitzer.caltech.edu/archanaly/contributed/browse.html.}, a
PSF simulator designed for {\it Spitzer} \citep{k02}.  An additional
background term was also included in the analysis, although this term
was statistically consistent with zero.

To streamline the processing, a few simplifications were made in the
modeling.  The galaxy center was fixed to match the observed galaxy
center.  The ratio of the minor and major axes for the disk and the
ring were treated as only one free parameter (i.e. the ratios were the
same for the ring and disk).  The position angle of the major axes for
the ring, disk, and bulge components was also treated as only one free
parameter.  The models were fit only to the region enclosed by the
ellipse where the B-band surface brightness reaches 25 mag
arcsec$^{-2}$, as defined in the Third Reference Catalogue of Bright
Galaxies \citep{ddcbpf91}.  The total of the flux densities for all the
components was forced to equal the total flux density within the
optical disk.  Uncertainties from the fits were generally estimated by
the variations in the parameters to fits performed on the north,
south, east, or west halves of the optical disk.

Because of variations among the different observations, the fits were
customized slightly for each waveband.  The customizations are
described below:

\begin{enumerate}
\item At 3.6-24~$\mu$m, matching the exact shape of the PSF of the nucleus
became too difficult (mainly because of the limitations of the
STinyTim software), so the inner $10^{\prime\prime}$ at 3.6-8~$\mu$m
and the inner $30^{\prime\prime}$ at 24~$\mu$m were not included in
the initial fits to the data (although the portion of the PSF that
extends outside this region from the nucleus was included).  After the
models were fit to the data, we used the results to determine the flux
densities of the bulge, disk, and ring components within this inner
region.  These components' flux densities were subtracted from the
total flux density within the central region to produce an estimate of
the nuclear flux densities that were next aperture corrected using the
PSFs from STinyTim.

\item At 3.6 and 4.5~$\mu$m, the ring feature was very faint.  If the ring
radius and width were allowed to vary as free parameters, the results
failed to model either the ring or disk components of the galaxy
properly.  Therefore, the ring radius and width were fixed to match
the parameters from the 5.7-70~$\mu$m data, although the ratio of the
minor and major axes and the position angle of the major axis
were allowed to vary.

\item At 70 and 160~$\mu$m, emission from the bulge is no longer
discernible in the data.  If the bulge is included in the 70~$\mu$m
model, the resulting bulge model produces physically implausible
results.  The bulge is therefore not included in these models.
Additionally, the value of the residual background term from the fits,
which is statistically equal to 0, demonstrates that the bulge
component does not need to be included.

\item In the 160~$\mu$m fits, the $40^{\prime\prime}$ resolution of
the data as well as the low sampling (where each location is only
observed 4 times) make the spatial parameters fit to the data
questionable.  Therefore, the parameters describing the shapes of the
profiles were fixed to values determined from the 5.7-70~$\mu$m data,
and the fits were performed only to determine the flux densities of
the different components.  Uncertainties were estimated by examining
the change in the fits when the spatial parameters were varied
$1\sigma$.
\end{enumerate}

As an example of the fits, the separate nucleus, inner disk, ring, and
bulge components of the model for the 24~$\mu$m data are shown in
Figure~\ref{f_modelparts}.  The observed 24~$\mu$m image, the model
24~$\mu$m image, and the residuals from subtracting the model from the
observed image are shown in Figure~\ref{f_modelcompare}.  Note that
the structures in the residual basically show substructure in the PSF
of the nucleus, which is a manifestation of the limitations of the
STinyTim software, and substructure in the ring, which is not
perfectly described by a Gaussian function and which does contain some
bright knots and faint extensions at the east and west ends of the
ring that possibly correspond to the faint outer ring described by
\citet{betal84} or faint spiral arm structures.  Otherwise, the
residual image shows no evidence of any additional structures present
in the data.

The parameters that describe the shapes of the model components are
presented in Table~\ref{t_fit_spatial}.  The weighted mean and
standard deviations of the 5.7-70~$\mu$m parameters describing the
disk and ring (used as described above in the 3.6, 4.5, and 160~$\mu$m
fits) are described in Table~\ref{t_fit_diskring}.  Note that, in the
5.7-70~$\mu$m range, the variation in the parameters between wavebands
is statistically small.  This demonstrates that the shape of the
features is relatively invariant across this wavelength range.  We
can infer that each component can be approximated as uniform in color
across this wavelength range, because if color gradients were present,
the parameters describing the shapes of the components would vary
across different wavebands.  For example, if the inner disk had a
color gradient, this would manifest itself as a variation in the scale
length of the exponential function that describes the inner disk, with
some wavebands having a shorter scale length than others.  Since no
statistical variation in this scale length is present, no color
gradient is present.  

The flux densities from the fits are given in Table~\ref{t_fit_fd}.
The uncertainties in the tables are from the fits; these uncertainties
effectively reflect the contribution of background noise or
substructures to the uncertainty.  Calibration uncertainties, which
are given in the final column of the table, are usually but not always
higher.

At 850~$\mu$m, the emission from outside the nucleus is negligible.
No significant structures are visible at above the $3\sigma$ level,
although some emission fainter than $3\sigma$ may be associated with
the dust ring.  As an approximation, however, this source can be
treated as a single unresolved source.  To obtain a nuclear flux
density, we simply measured the emission within the central $1^\prime$
of the galaxy.  The measured flux density is 0.25~Jy with a
calibration uncertainty of 10\%.  Note that this flux density
measurement is consistent with the 870~$\mu$m flux density measurement
of 0.230 $\pm$ 0.035 Jy measured by \citet{kwd05}.  However, we note
that some fraction of the flux density within this aperture may
originate from the inner disk and ring.  We used the parameters given
in Tables~\ref{t_fit_diskring} and \ref{t_fit_fd} as well as the
models of the SEDs of each component discussed in
Section~\ref{s_sedanalysis} to construct models of the disk and ring
at 850~$\mu$m.  These models show that that 20\% or less of the
measured 850~$\mu$m emission may originate from the inner disk and
ring.  This estimate, however, relies on an extrapolation of SED
models from shorter wavelengths; the 850~$\mu$m flux densities of
these sources cannot be constrained with these data.  (Note that the
predicted 850~$\mu$m surface brightnesses of the disk and ring are
equivalent to the noise levels in the map.)  Therefore, in
Table~\ref{t_fit_fd}, we report the flux density of the nucleus to be
0.25~Jy with an uncertainty of 0.05~Jy (20\%).

\subsubsection{Analysis of the Spectral Energy Distributions of the
Separate Model Components \label{s_sedanalysis}}

The SEDs of the total emission within the optical disk of the galaxy
as well as the SEDs for the nucleus, disk, ring, and bulge components
are presented in Figure~\ref{f_sed}.  For comparison to the SEDs of
nearby galaxies, we used the results of fitting semi-empirical dust
models to the global SEDs of SINGS galaxies in \citep{detal05} to
determine what the SED of a typical galaxy was.  The median $\alpha$
(the index for the power law that describes the distribution of the
intensities of the raditation fields that heat the dust) of the models
fit to these galaxies is 2.375.  The semi-empirical model with this
$\alpha$ is plotted on top of the SEDs in Figure~\ref{f_sed} as dotted
lines.  Additionally, emission models comprised of starlight added to
the physical dust models of \citet{ld01} have been fit to the SEDs of
the nucleus, the inner disk, and the ring.  The models treat the dust
as a mixture of amorphous silicate and carbonaceous grains (including
PAHs) with a size dsitribution that reproduces the optical and
ultraviolet extinction in the Milky Way.  The dust grains in this
model are then heated by starlight with a distribution of intensities
given by a power law.  In fitting this model to the data, the
parameters allowed to vary included the mass of the dust, the index
for the power law describing the distribution of intensities of the
raditation fields that heat the dust, the maximum and minimum of the
radiation fields, the PAH abundance, and the starlight intensity.
These models are plotted on top of the SEDs in Figure~\ref{f_sed} as
solid lines.

The global SED appears similar to the typical SEDs of nearby spiral
galaxies \citep[e.g.][]{rtbetal04, detal05}.  The SEDs of the
individual components, however, look very different from each other.
In the following paragraphs, the SEDs of the individual components are
discussed as they appear in Figure~\ref{f_sed} except for the SED of
the nucleus, which is discussed last.

The disk 70~$\mu$m~/~160~$\mu$m color is relatively warm compared to
the ring emission and the total emission but it is close to the what
is expected for the typical galaxy in \citet{detal05}.  The
far-infrared color temperature (defined in this paper as the
temperature of the blackbody function modified by a $\lambda^{-2}$
emissivity law that fits the 70 and 160~$\mu$m data)can be used as an
approximation of the dust temperature.  In the case of the inner disk,
the far-infrared color temperature is $26\pm2$~K.  The interstellar
radiation field determined from fitting the \citet{ld01} model to the
data can also be used as an indicator of how much the dust is heated.
In the inner disk, the average intensity of the radiation field is
approximately four times the local value (i.e. the interstellar
radiation field near the Sun).  The \citet{ld01} models also predict
that the total dust mass is $3.5\times10^5$~M$_\odot$.  Note that this
is a relatively small amount of dust compared to the total dust mass
of the ring (given in the next paragraph).

The colors of the dust ring are comparable to the colors of typical
nearby galaxies, although the 160~$\mu$m emission is slightly higher
than what is expected for the typical SED in \citet{detal05}.  The
far-infrared color temperature is $19 \pm 2$~K, and the average
intensity of the radiation field determined from the \citet{ld01}
model fits is equivalent to the local value.  The total dust mass
predicted for the ring by the model is $7.9\times10^6$~M$_\odot$.
Assuming that the gas-to-dust mass ratio is similar to the value of
165 for the Milky Way \citep{l05}, the gas mass as determined from the
dust mass is $1.3\times10^{9}$~M$_\odot$.  For comparison, the atomic
gas mass given in \citet{betal84} is $3.2\times10^{8}$~M$_\odot$ (when
scaled to a distance of 9.2~Mpc) and the upper limit of the molecular
gas mass given in \citet{betal91} is given as
$4.4\times10^{8}$~M$_\odot$ (when scaled to a distance of 9.2~Mpc).
The upper limit on the total atomic and molecular gas mass is
therefore approximately $7.6\times10^{8}$~M$_\odot$, which is within a
factor of 2 of the gas mass estimate of the dust mass.  These results
suggest that the majority of the expected mass of dust in this galaxy
can be accounted for in the ring.

The SED of the bulge decreases monotonically from short wavelengths to
long wavelengths and virtually disappears at 70~$\mu$m. The slope of
the SED is shallower than what is expected from blackbody emission,
possibly hinting at the presence of hot dust in the atmospheres of
evolved stars in the bulge \citep[see][for example]{bgs98}.  At
8~$\mu$m, half of the global flux from NGC~4594 originates from the
bulge, and at 24~$\mu$m, the bulge still contributes 1/3 of the total
flux.

Of most interest, however, is the unusual SED of the nucleus.  Note
the relatively hot colors compared to the ring and disk.  The
24~$\mu$m~/~70~$\mu$m and 24~$\mu$m~/~160~$\mu$m colors are relatively
high.  The far-infrared color temperature is $25 \pm 2$~K, and
according to the \citet{ld01} dust model fit to the data, the average
radiation field in the nucleus is approximately six times the local
value.  Strangely, the 70 - 850~$\mu$m emission almost appears to be
flat, in contrast to the ring and inner disk SEDs.  Of particular
interest is the 160 and 850~$\mu$m emission.  Placed in the context of
the SEDs of other nearby galaxies, the 70~$\mu$m emission and
especially the 160~$\mu$m emission is relatively weak compared to
emission at 24~$\mu$m and shorter wavelengths.  The 850~$\mu$m
emission is of interest in that it is abnormally high compared to the
70 and 160~$\mu$m data and it cannot be reproduced by the \citet{ld01}
dust model fits.

Note that the nuclear emission represents the enhancement of emission
by the AGN itself.  As demonstrated in Section~\ref{s_spectra}, star
formation is either absent from the nucleus or makes only a negligible
contribution to the radiation field in the center of the galaxy.
Passive heating from evolved stars in the inner disk or the bulge is
another possibility.  However, note that the scale lengths of the
inner disk and the bulge do not vary between 5.7 and 70~$\mu$m.  This
indicates that the infrared colors of the bulge and disk do not vary
and that the nuclear emission is not the result of an enhancement in
the nuclear infrared colors from evolved stars.  The nuclear infrared
emission must be enhanced by the AGN.

\section{Discussion \label{s_discussion}}

The 24 - 850~$\mu$m regime of the nuclear SED leads to two significant
results.  The first result is the especially low 160~$\mu$m emission
from the nuclear region (compared to emission at 24~$\mu$m and shorter
wavelengths) and its possible connection to the weak circumnuclear
star formation activity (as revealed by the absence of PAH emission).
The other result is the unusually high 850~$\mu$m emission, which
appears to originate from a source other than the $\sim20-30$~K dust
typically found in the interstellar medium of other galaxies or in the
ring and inner disk of this galaxy.  We discuss these two results
below.

\subsection{The Implications of the Weak Nuclear 160~$\mu$m Emission}

The weak 160~$\mu$m emission from large, cool grains in the
environment around this AGN is best understood when placed in the
context of the relation between far-infrared emission and either AGN
or star formation activity in other AGN host galaxies.  Early results
from IRAS, such as \citet{rrj87} and \citet{rc89}, had suggested that
far-infrared emission from nearby Seyert galaxies was associated with
star formation, not AGN activity.  In later studies of Seyfert
galaxies, including \citet{metal95}, \citet{rp97}, \citet{prs98}, and
\citet{prf00}, the far-infrared dust emission, while still observed to
be strongly peaked near the centers of the galaxies, had been shown to
be associated with star formation, and the contribution of dust heated
by the AGN to the nuclear emission was minimal.

In contrast to the observations of Seyfert galaxies cited above, the
160~$\mu$m emission commonly associated with star formation or cirrus
emission does not peak in the center of NGC~4594.  PAH emission is
also largely absent from inside the ring, which is consistent with the
AGN being the dominant energy source in this region.  The molecular
gas that is associated with star formation seems to be limited to the
dust ring. CO data presented in \citet{betal91} show that the CO
emission detected from the galaxy corresponds to only locations in the
dust ring, not the nucleus.  CO observations in \citet{yetal95} show
that the molecular gas is uniformly distributed along the major axis,
which implies that it corresponds best to the dust emission from the
ring (although note that the individual pointings only detect CO at
the $2\sigma$ level).  The interstellar gas around the AGN appears to
be predominantly hot X-ray emitting gas \citep{pffta02}, which cannot
fuel star formation and which may be too low in density to fuel
enhanced AGN activity.

The differences between NGC~4594 and Seyfert galaxies in terms of
circumnuclear far-infrared emission therefore appears to be connected
to circumnuclear star formation activity.  In Seyfert galaxies, star
formation and the far-infrared dust emission associated with it are
typically found near the AGN.  The presence of strong star formation
(or relatively young stellar systems) near Seyfert nuclei has been
noted previously \citep{gp93, metal95, oetal95, hetal97, ghlmkwkk98,
oetal99, ghl01}, although the observations have generally found the
presence of strong circumnuclear star formation in Seyfert 2 galaxies
rather than Seyfert 1 galaxies.  In the case of the AGN in NGC~4594,
however, the cool gas that fuels both enhanced star formation and AGN
activity is not present.  Therefore, the region is devoid of the
long-wavelength far-infrared emission from the $\sim20$~K dust
associated with molecular gas.

The absence of recent star formation is observed in at least a
significant fraction of nearby LINER and low luminosity AGN nuclei
\citep{letal98, cetal04, getal04, bj04} and the relative lack of
mid-infrared dust emission in LINERs compared to other galaxies
\citep{betal02} has been observed before.  The data taken for all
SINGS galaxies as well as other mid- and far-infrared surveys of
nearby LINERs and low luminosity AGN should be used to determine
whether AGN nuclei like NGC~4594 are also weak yet hot infrared
sources.

\subsection{The Origin of the 850 $\mu$m Emission \label{s_850}}

In most galaxies, the 850~$\mu$m waveband is dominated by
$\sim$20-30~K dust emission \citep[e.g.][]{detal00, de01, betal03,
rtbetal04}.  In NGC~4594, however, the difference in morphology
between the 160~$\mu$m and 850~$\mu$m images as well as the relatively
high ratio of 850~$\mu$m to 160~$\mu$m flux densities for the nucleus
clearly demonstrate that the 850~$\mu$m emission is not from
$\sim$20-30~K dust.  The 850~$\mu$m emission must originate from
another source, possibly one connected to the high-energy phenomena
observed in other wavebands.  We examine a number of alternative
850~$\mu$m emission mechanisms below.

\subsubsection{Very Cold Dust or Exotic Dust Emission}

When excess emission is observed at submillimeter wavelengths compared
to the modified blackbody emission at far-infrared wavelengths, one
common explanation is that the excess emission comes from very cold
dust at temperatures in the 5 - 10~K range.  Therefore, we will
consider whether the excess submillimeter emission from the nucleus of
NGC~4594 is from such dust emission.

Qualitatively, it seems unlikely that a very cold dust component could
be responsible for the excess emission seen at 850~$\mu$m.  First,
Figures~\ref{f_images} and \ref{f_sed} show that the dust in the
nucleus is strongly heated by the AGN.  The high temperatures of the
circumnuclear environment are also evident in the H$\alpha$ emission
\citep{hfs97}, which demonstrate that photoionizing photons are
present, and in the X-ray observations \citep{pffta02, pbfk03}, which
demonstrate that both hot X-ray gas and hard X-ray synchrotron
emission is present.  For a large mass of very cold dust to exist in
such an environment seems unlikely.  Furthermore, to have a
significant very cold dust component present without any strong
emission from a $\sim$20-30~K dust component seems unlikely.

To examine this further, we calculated the minimum dust mass that
would be needed to produce the 850~$\mu$m emission using the equation

\begin{equation}
M_{dust} = \frac{D^2 f_{850\mu m}}{\kappa_{850\mu m} B(T)_{850\mu m}}
\end{equation}

\noindent where $D$ is the distance to the object (9.2~Mpc),
$f_{850\mu m}$ is the 850~$\mu$m flux density, $\kappa_{850\mu m}$
represents the absorption opacity of the dust at 850~$\mu$m
\citep[0.431~cm$^2$~g$^{-1}$;][]{ld01}, and $B(T)_{850\mu m}$ is the
surface brightness of a pure blackbody at a temperature $T$.  The
minimum dust mass will correspond to the warmest temperature that is
still consistent with the data.  This temperature will describe the
modified blackbody function that fits both the 160~$\mu$m and
850~$\mu$m measurements.  Using an emissivity varying as
$\lambda^{-2}$ (which approximates the emissivity of dust in the
far-infrared and submillimeter; see \citealt{ld01}), the modified
blackbody that best fits the 160 and 850~$\mu$m data has a temperature
of $9 \pm 2$~K.  This is consistent with a dust mass of
$1.9\times10^7$~M$_\odot$.  Note, however, that a substantial part of
the 160~$\mu$m emission probably comes from warmer dust emission, as
would be implied from the emission at 70~$\mu$m (see, for example, the
model fit to the nuclear SED in Figure~\ref{f_sed}).  The temperature
of this hypothetical very cold component could be lower, which would
drive the dust mass of the very cold component higher.

Assuming a gas-to-dust mass ratio of 165, the minimum dust mass
calculated above implies a total gas mass of
$3.2\times10^{9}$~M$_\odot$ in the central $15^{\prime\prime}$ region
of the galaxy (or within a radius of 330 pc of the center).  This gas
mass is higher than the upper limit of $7.8\times10^{8}$~M$_\odot$ for
the global total of atomic and molecular gas in NGC~4594 (as
calculated in Section~\ref{s_sedanalysis}), so it is implausible to
expect so much gas to be present in the center.  Furthermore, the
black hole in the AGN itself is $10^9$~M$_\odot$.  It is implausible
to think that a mass of cold gas multiple times larger than the mass
of the central AGN could be located in an environment where
significant numbers of high energy photons from the AGN and hot gas
are present \citep[e.g.][]{pffta02, pbfk03}.  Even if only half the
submillimeter emission originates from $\sim10$~K dust, the implied
molecular gas mass is still larger than the mass of the black hole.
Therefore, we reject the possibility that a very cold dust component
is generating the 850~$\mu$m emission.

Another possibility is that the submillimeter emission might be
produced by grains with a much larger submillimeter opacity than the
grains that appear to be responsible for the bulk of the far-infrared
and submillimeter emission observed from most galaxies.  Exotic grains
such as fractal grains have been proposed as an explanation for the
excess submillimeter emission observed from objects such as the Milky
Way \citep{retal95} and NGC~4631 \citep{dkw04}.  Such dust would have
temperatures of 5 - 10~K and would radiate predominantly at
submillimeter wavelengths.

This explanation, however, has problems.  Even if the exotic grains
have a sufficiently high ratio of submillimeter opacity to
optical-ultraviolet opacity to remain at $\sim$10K in the intense
radiation field, the required dust mass would remain unacceptably
large unless the grains have a submillimeter opacity orders of
magnitude larger than normal dust.  While very large submillimeter
opacities have been reported for some laboratory materials
\citep[e.g., material "BE" of][]{metal98}, no evidence suggests that
such materials exist in interstellar space (e.g., B. T. Draine 2006,
in press).

\subsubsection{CO Emission}

The contribution of CO(3-2) emission to broadband 850~$\mu$m emission
has been a concern when using these wavebands for measurements of dust
SEDs in the 15 - 30~K range.  Usually, CO(3-2) emission is only a
minor contribution to the total submillimeter emission, although some
exceptional cases have been identified \citep[e.g.][]{ketal01}.

\citet{betal91} detected the CO(1-0) line $140^{\prime\prime}$ east
and west of the nucleus in NGC~4594 but did not detect the central
position because of instrumental problems.  The data in
\citep{yetal95} show that the CO(1-0) emission does not peak in the
center of the galaxy but is instead uniformly distributed along the
major axis (although note that CO is detected in the individual
pointings at only the $2\sigma$ level).  Qualitatively, these data
suggest that the CO(1-0) emission is primarily in the dust ring, not
in the nucleus.

To estimate the possible 850~$\mu$m flux density of the CO(3-2) line
in the nucleus, we will use the \citep{yetal95} CO(1-0) upper limit of
1.6~K~km~s$^{-1}$ for the center of NGC~4594.  Using the conversion
factor of 42~Jy/K from Young et al. and the assumed line width of
400~km~s$^{-1}$, the upper limit of the flux density of the CO(1-0)
transition is $\sim0.017$~Jy. Assuming that the CO J levels are
thermalized at a rotational temperature T$_{rot} >$15~K, we estimate
an upper limit for the flux density of $\sim1.5$~Jy for the CO(3-2)
transition. Using the relative width of the potential line
($\sim400$~km~s$^{-1}$) and the SCUBA 850W filter bandpass (40~GHz) we
obtain an upper limit for the contribution of the CO flux density at
850~$\mu$m of $\sim20$~mJy, which is much less than the 250~mJy
measured for the nucleus.  We therefore conclude that emission from CO
cannot explain all of the emission at 850~$\mu$m.

\subsubsection{Synchrotron Emission}

Synchrotron emission has been observed at submillimeter wavelengths in
such AGN as NGC~4374 \citep{lsr00} and NGC~1275 \citep{isb01}.
Therefore, it is a natural possibility as the source of the
submillimeter emission in this galaxy.

To determine the contribution of synchrotron emission at 850~$\mu$m,
we will extrapolate from radio observations to the submillimeter.  The
best published multiwavelength radio data taken for this galaxy come
from \citet{detal76} and \citet{hvd84}.  Their results show that all
of the radio emission from NGC~4594 originates from the AGN, which
appears as an unresolved source.  The SED shows that the radio
emission consists of synchrotron emission with a break at
approximately 6~cm (5~GHz) caused by synchrotron self-absorption.  A
power law fit to the 2-6~cm (15.0-5.0~GHz) data from \citet{hvd84}
shows that flux density is proportional to $\lambda^{0.2}$.  Using
this power law fit to extrapolate from centimeter to submillimeter
wavelengths as shown in Figure~\ref{f_sed_agn}, we predict an
850~$\mu$m flux density of 0.056~Jy, which is approximately a factor
of 4 lower than the measured flux density.

Variable submillimeter synchrotron emission is a possibility.  The
synchrotron emission from the AGN is variable, but the variability may
not be significant enough to explain the difference between the
observed 850~$\mu$m flux density and the extrapolations of synchrotron
emission to those wavelengths.  Decades of observations show that the
6~cm radio emission has only been observed to vary at the 10 - 20~\%
level \citep{detal76, efm83, kwd05}.  \citet{betal88} and references
therein found that the 20~cm flux density increased by $\sim70$~\%
between 1980 and 1985.  However, this increase was not observed at
other wavelengths, nor is it clear that this variability is seen at
other wavelengths.  At 3.6~cm, \citet{tpkbo00} measured a flux density
of 84.7 $\pm$ 0.05 ~mJy in 1995-1996, and \citet{kwd05} measured a
flux density of 90 $\pm$ 10~mJy in 2003.  These values are lower than
the 113 $\pm$ 12~mJy measured by \citep{detal76}.  However, it is
unclear whether this represents a decrease in the luminosity of the
source itself or an improvement in the calibration of measurements at
these wavelengths.  Moreover, this decrease is unlikely to correspond
to an increase in synchrotron emission at 850~$\mu$m unless the power
law describing the synchrotron emission changes as well.  Finally, we
would like to state that the consistency between the 850~$\mu$m flux
density measured in this paper and the 870~$\mu$m flux density
measured in \citet{kwd05} implies that the submillimeter emission is
not variable at a level greater than the measurement uncertainties
(10~\%), at least on periods of $\sim3$~yr.  Therefore, it is unlikely
that the submillimeter emission observed at 850~$\mu$m was captured at
a maximum in emission while the radio emission was taken at a minimum.

Nonetheless, we are still hesitant to rule out synchrotron emission
entirely as a source of the 850~$\mu$m emission.  The extrapolation
from radio wavelengths did yield an estimate of the 850~$\mu$m
emission that was within less than a factor of 10 of the measured flux
density.  The power law used in the extrapolation was determined using
only three data points, and the extrapolation extended over a factor
of 10 in wavelength.  So, we caution that the extrapolations are not
entirely reliable and that the actual synchrotron emission may be
higher or lower than what we determined here.

It is also possible that the 850~$\mu$m emission is from synchrotron
emission unrelated to the centimeter wavelength emission.  Such
synchrotron emission would need to be self-absorbed at wavelengths
longer than 850~$\mu$m.  To determine if such a synchrotron
component is present would require further observations in multiple
wavebands between 350~$\mu$m and 2~cm.

\subsubsection{Bremsstrahlung}

A final possibility is that bremsstrahlung emission is responsible for
the 850~$\mu$m emission.  Bremsstrahlung emission has been detected in
some nearby objects, such as M~82 \citep[e.g.][]{c92}, but has never
been identified as a dominant source at 850~$\mu$m.

Since bremsstrahlung emission and recombination line emission both
originate from ionized hydrogen gas, it is possible to relate the two
emission processes to infer the expected bremsstrahlung emission.  To
convert the H$\alpha$ flux to 850~$\mu$m flux density, we will use the
equations from Appendix A in \citet{cd86}.  We assume that the
temperature is $\sim10^4$~K and the contribution of He{\small II} to
the bremsstrahlung emission is negligible.  This gives the conversion

\begin{equation}
f_\nu(850 \mu m; Jy) = 5.29 \times 10^9 f(H\alpha; erg \ cm^{-2} \ s^{-1}) 
\end{equation}

\citet{hfs97} give the H$\alpha$ flux as as measured in a
$2^{\prime\prime}$ wide region as $1.05 \times
10^{-13}$~erg~cm$^{-2}$~s$^{-1}$.  This yields an expected 850~$\mu$m
bremsstrahlung flux density of $5.5 \times 10^{-4}$ Jy.  This is a
factor of 400 too low to explain the observed 850~$\mu$m flux density
from the AGN. Note that the Ho et al. measurements are not corrected
for dust extinction.  However, the low extinction measured with the
H$\alpha$~/~H$\beta$ ratio in Ho et al. implies that the extinction
correction for the H$\alpha$ flux would not significantly change the
estimate of the 850~$\mu$m bremsstrahlung emission.  Therefore, we
conclude that some other emission process must be responsible.

\subsubsection{Conclusions on the Origin of the 850~$\mu$m Emission}

This analysis has shown that no known emission mechanism can
satisfactorily explain the observed 850~$\mu$m flux density of 0.25~Jy
from the nucleus of NGC~4594.  Synchrotron emission seemed to be the
most plausible single mechanism to explain the 850~$\mu$m emission,
although the 0.056~Jy flux density at 850~$\mu$m inferred from the
radio synchrotron emission falls short of the observed flux density.
However, it it still possible that this synchrotron emission
contributes a fraction of the total 850~$\mu$m flux density.  CO
emission, with a flux density of 0.020~Jy, and bremsstrahlung
emission, with a flux density of $5.5 \times 10^{-4}$ Jy, are also
physically plausible emission mechanisms that may contribute to the
850~$\mu$m band.  Very cold dust (at temperatures of $\sim10$~K) is
not a physically plausible source of 850~$\mu$m emission, although
some small amount of the flux density (approximately 2\% according to
the \citet{ld01} dust models fit to the nuclear SED in
Section~\ref{s_sedanalysis}) may be dust emission from warmer dust.

Even though multiple emission mechanisms may contribute to the total
nuclear emission at 850~$\mu$m, the total flux density of
$\sim0.080$~Jy from all known physically plausible emission mechanisms
(synchrotron, CO, bremsstrahlung, and warm dust emission) still falls
short of the 0.250~Jy flux density observed at 850~$\mu$m.  The
possibility remains that some mechanism that is dissociated from the
emission mechanisms observed in all other wavebands, such as
synchrotron emission that is self-absorbed longward of 850~$\mu$m or
unidentified spectral line emission, could be responsible for a
significant fraction of the 850~$\mu$m nuclear emission.  Additional
submillimeter and millimeter photometry and spectroscopy are needed to
identify the source of the 850~$\mu$m emission, which should provide
further clues about the nature of this and other similar massive AGN.

\section{Conclusions}

We have extracted the SEDs of the nucleus, inner disk, ring, and bulge
in the Sombrero Galaxy, NGC~4594.  The SED of the nucleus is notably
unusual in that the emission spectrum requires the dust to be hot,
that the 160~$\mu$m emission is notably weak compared to
emission at 24~$\mu$m and shorter wavelengths, and that some source
other than large $\sim$20-30~K dust grains powers the 850~$\mu$m
emission.

The relatively weak 160~$\mu$m nuclear emission appears to indicate
that the LINER activity seen in this galaxy is a result of the lack of
cool gas needed to fuel stronger Seyfert activity.  The corresponding
lack of star formation implied by the data is in line with previous
studies, which have found weak nuclear star formation activity in many
(but not all) LINERs and low luminosity AGN.  Moreover, the strong
far-infrared dust emission from the ISM near Seyfert nuclei as well as
the strong circumnuclear star formation in Seyfert galaxies also
provide credence to this idea.

The abnormally high 850~$\mu$m emission from the nuclear region cannot
be explained as any kind of dust emission, as synchrotron emission
related to that seen at centimeter wavelengths, or as bremsstrahlung
emission.  A combination of mechanisms may be necessary to explain the
850~$\mu$m emission.  It is possible that the 850~$\mu$m emission may
come from an emission source that cannot be inferred from observations
in other wavebands, such as synchrotron emission that is self-absorbed
at wavelengths longer than 850~$\mu$m or unidentified spectral line
emission in the 850~$\mu$m band.  Further observations at
submillimeter and millimeter wavelengths are needed to determine the
nature of the emission.

Future observations of low luminosity AGN with the {\it Spitzer Space
Telescope} may reveal more galaxies with similarly weak nuclear
160~$\mu$m emission.  Already, SINGS observations of the
low-luminosity AGN NGC~2841 indicate that it may have a similar
spectral energy distribution.  Such anomalous nuclear emission should
be placed into context by comparing these sources to similar LINERs
and to Seyfert galaxies.  The end result may be the identification of
key differences between these two classes of objects.  We also
anticipate that a comparison of 160~$\mu$m {\it Spitzer} images with
15$^{\prime\prime}$ resolution submillimeter or millimeter data will
reveal other galaxies with anomalously high submillimeter or
millimeter emission, which should lead to identification of the source
of the emission and a more complete view of the energetics of AGN.

\acknowledgments 
Support for this work, part of the {\it Spitzer Space Telescope}
Legacy Science Program, was provided by NASA through contract 1224769
issued by the Jet Propulsion Laboratory, California Institute of
Technology under NASA contract 1407.  BTD was supported in part by NSF
grant AST-0406833.

\clearpage

\begin{figure}
\center{\includegraphics[height=8in]{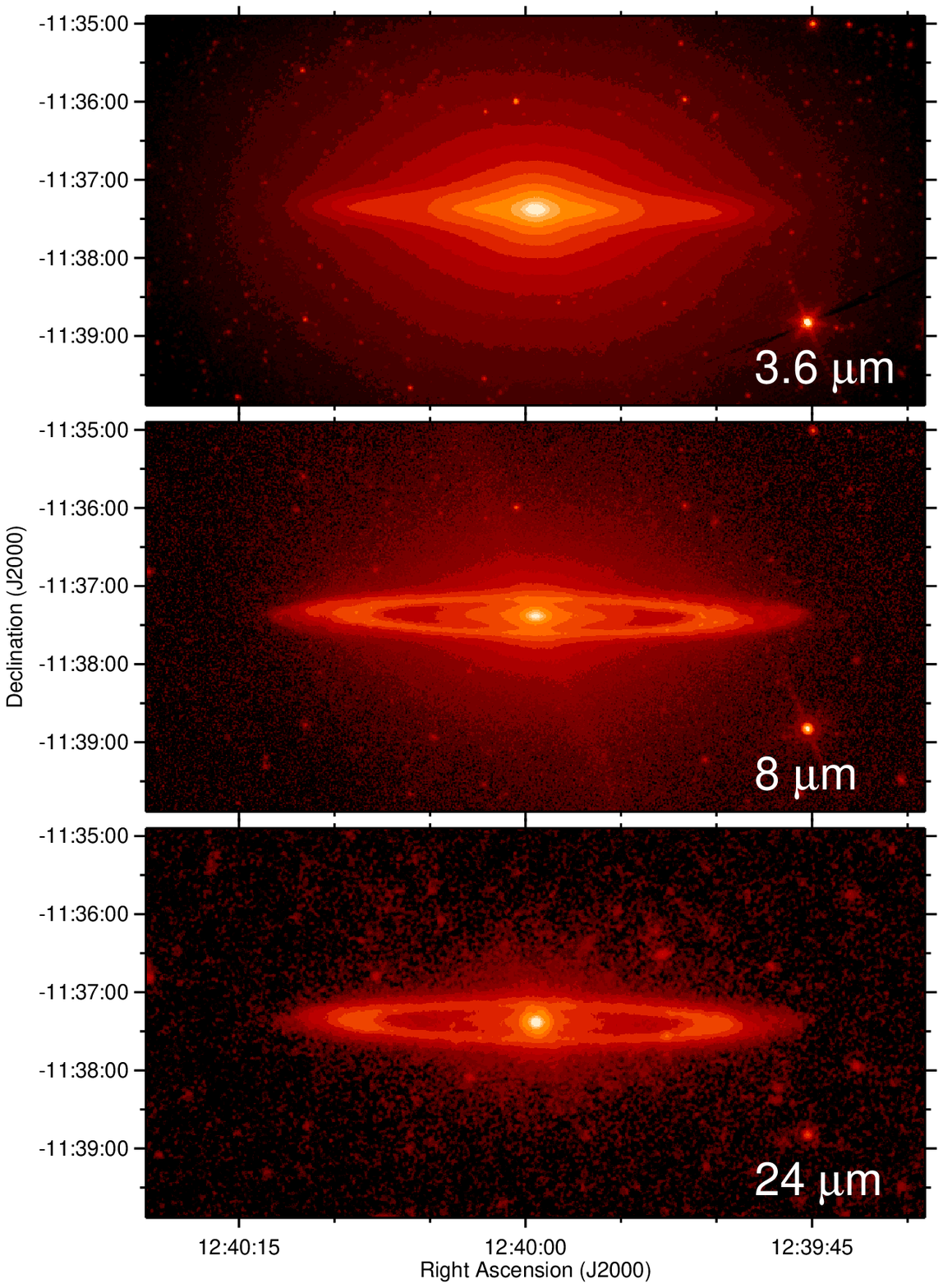}}
\caption{3.6, 8, 24, 70, 160, and 850~$\mu$m images of NGC~4594.  Each
images is $10^\prime \times 5^\prime$, with north up and east to the
left.  The scaling of the brightness in all images is logarithmic.
Note that only the inner $\sim3^\prime \times \sim2.25^\prime$ region
was covered in the 850~$\mu$m band and that the very bright and dark
pixels at the edges of this region are an artifact of the data
processing.}
\label{f_images}
\end{figure}

\clearpage

\addtocounter{figure}{-1}
\begin{figure}
\center{\includegraphics[height=8in]{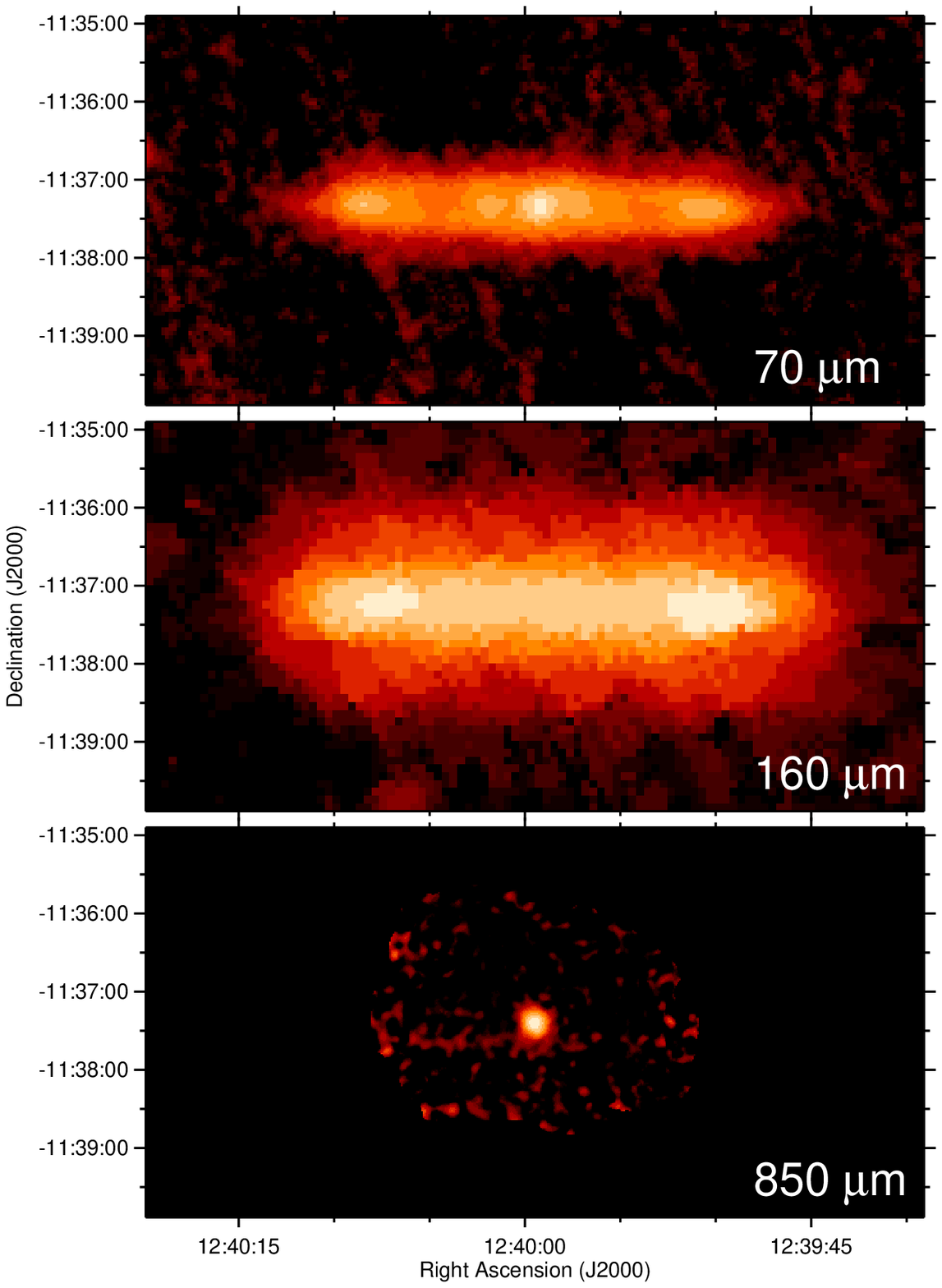}}
\caption{Continued.}
\end{figure}

\clearpage

\begin{figure}
\plotone{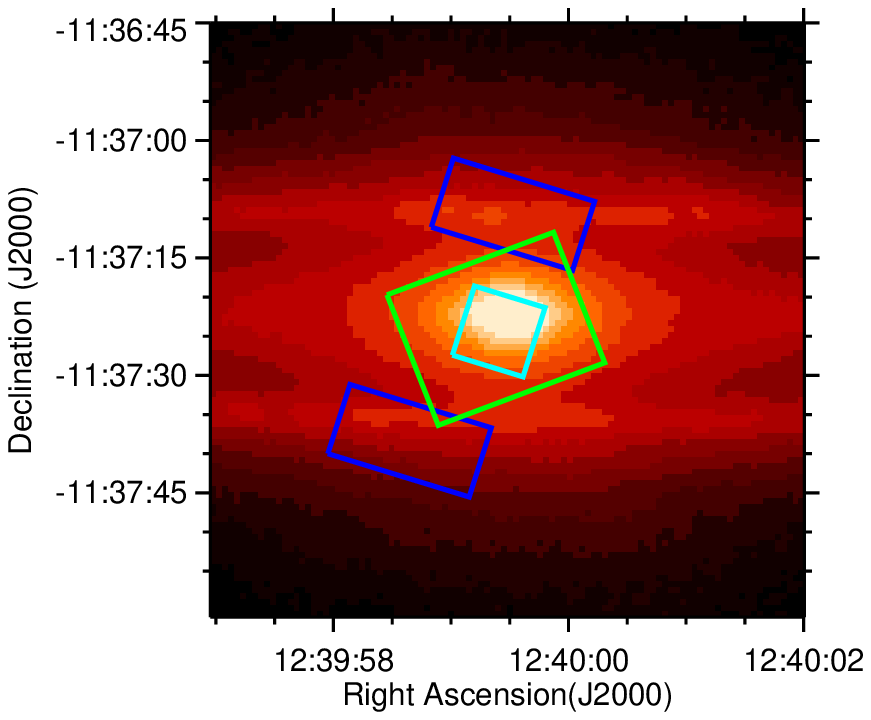}
\caption{The 8~$\mu$m image of the central $75^{\prime\prime}$ of
NGC~4594, with boxes showing the regions in which mid-infrared spectra
were extracted.  The cyan square in the center shows where the
low-resolution 5-38~$\mu$m spectrum for the nucleus was extracted.
The dark blue rectangles at top and bottom show where the
low-resolution 5-38~$\mu$m spectrum for the ring was extracted.  The
green rectangle in the center shows where the high-resolution
25-27~$\mu$m spectrum for the nucleus was extracted.}
\label{f_extregions}
\end{figure}

\clearpage

\begin{figure}
\plotone{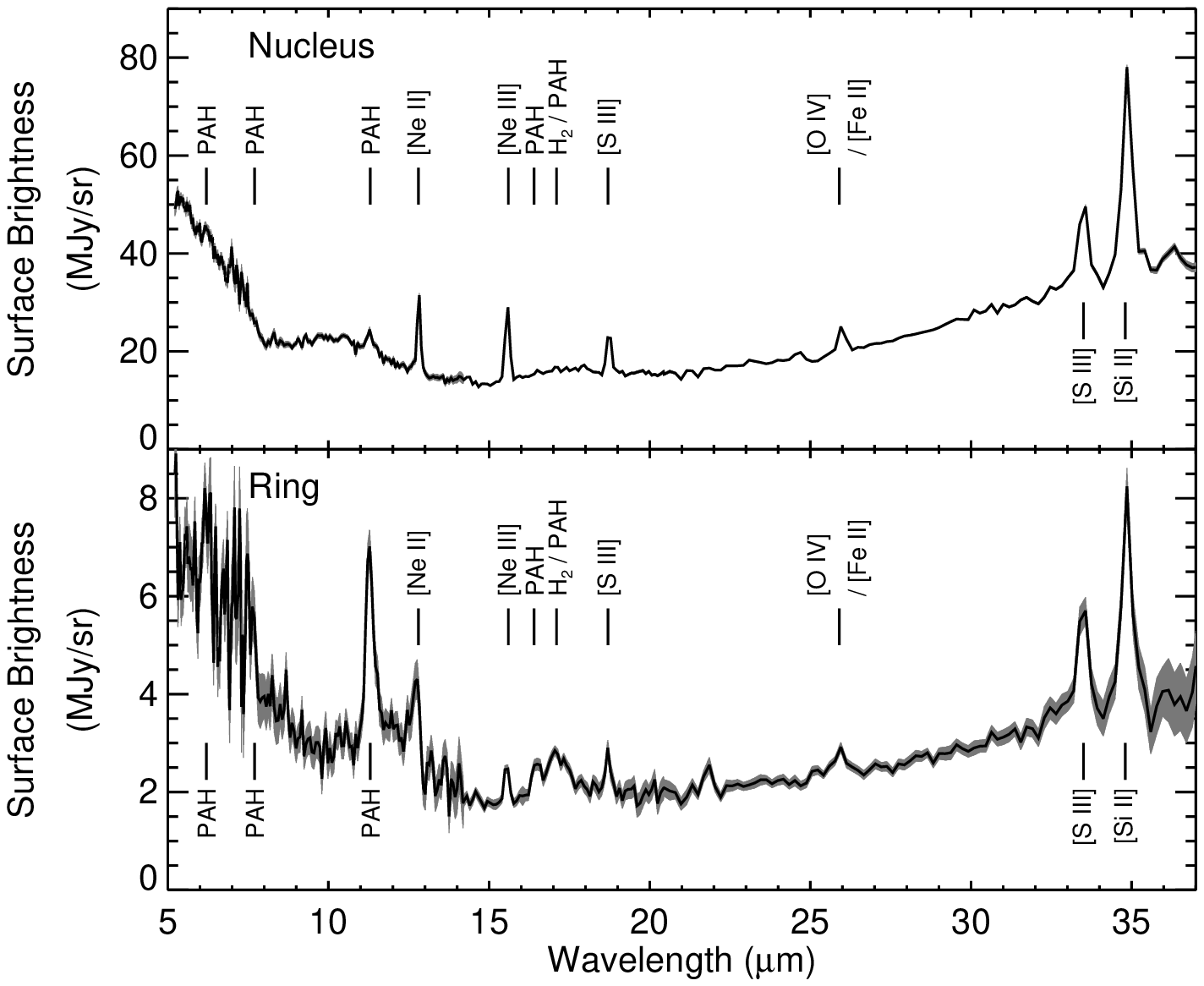}
\caption{5-38~$\mu$m low-resolution IRS spectra of the nucleus and the
ring in NGC 4594.  The gray regions around the lines represent the
uncertainties in the spectra.  Major spectral features are identified
in the plot.  Note the weak or absent PAH features in the nucleus, in
contrast to the strong PAH features in the ring.  Because of the
limitations of the spatial resolution of the telescope, the spectral
features longward of 20~$\mu$m in the ring spectrum probably include
emission from the center.  See the text for additional details.  See
Figure~\ref{f_spectra_o4} to see a higher-resolution 25.5-26.5~$\mu$m
spectrum of the nucleus where the [O~{\small IV}] and [Fe~{\small II}]
lines are separated.}
\label{f_spectra}
\end{figure}

\clearpage

\begin{figure}
\begin{center}
\includegraphics{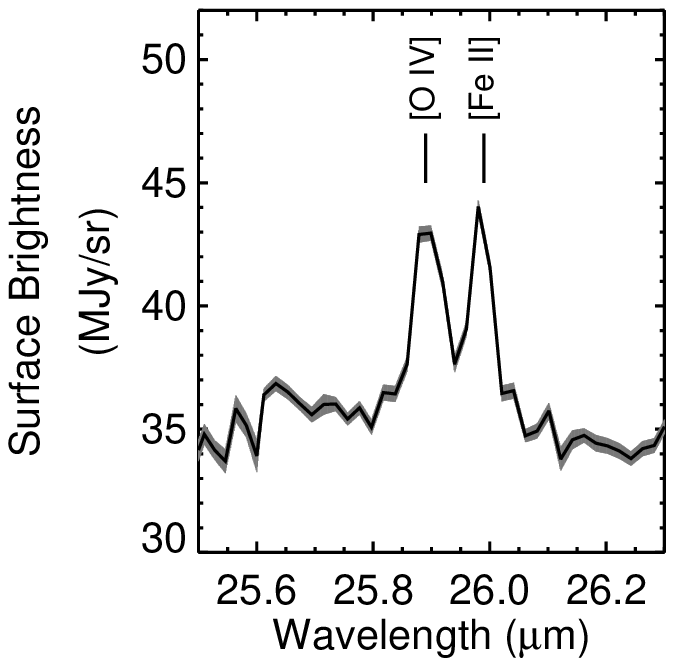} 
\end{center}
\caption{25.5-26.5~$\mu$m high-resolution IRS spectrum of the nucleus
in NGC 4594. The gray regions around the lines represent the uncertainties in
the spectra.  This plot shows the detail in the [O~{\small IV}] and
[Fe~{\small II}] line emission.}
\label{f_spectra_o4}
\end{figure}

\clearpage

\begin{figure}
\plotone{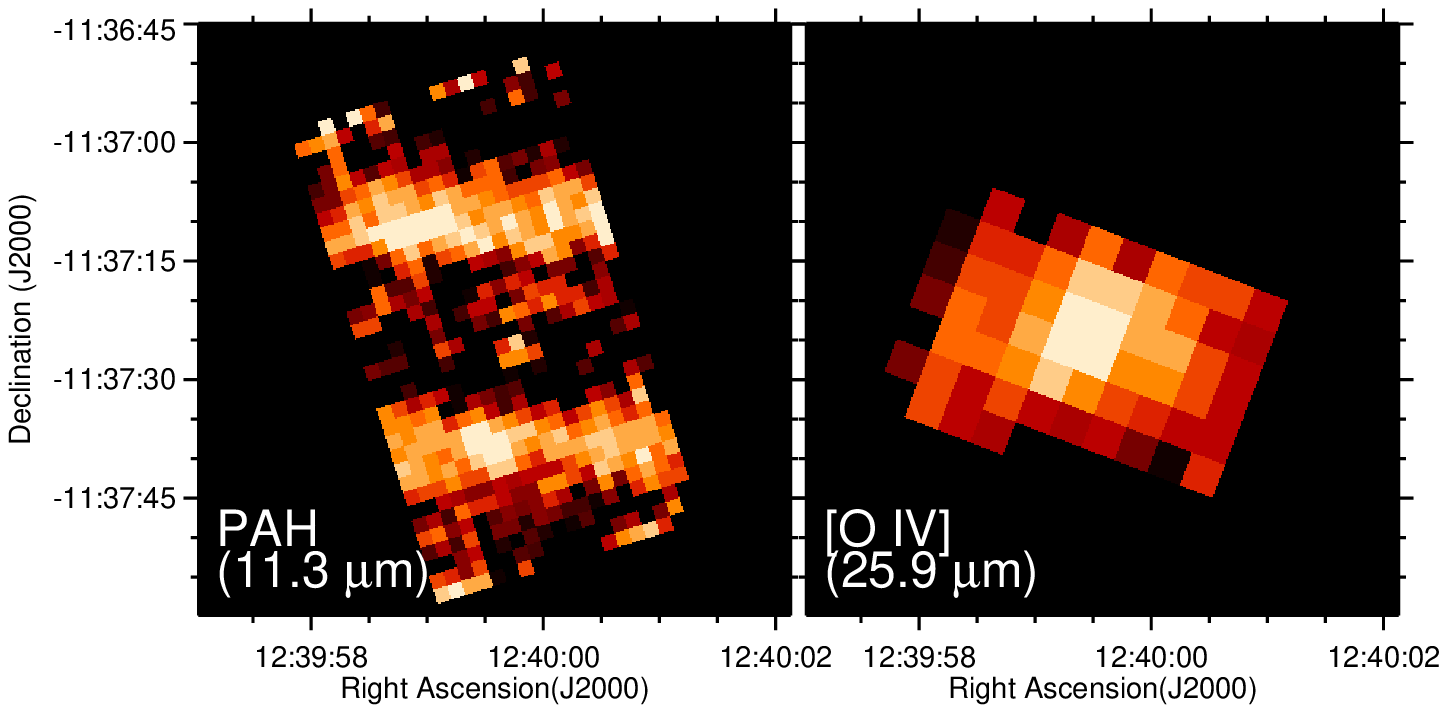}
\caption{Images of the 11.3~$\mu$m PAH feature (left) and the 25.9
[O~{\small IV}] line (right) for the inner $75^{\prime\prime}$ of
NGC~4594.  These are images made from the IRS spectral cubes.  The
continuum has been identified and subtracted from the two wavebands on
a pixel-by-pixel basis.  Note that the PAH emission originates
primarily from the ring, whereas the [O~{\small IV}] emission
originates primarily from the nucleus.}
\label{f_imgline}
\end{figure}

\clearpage

\begin{figure}
\center{\includegraphics[height=8in]{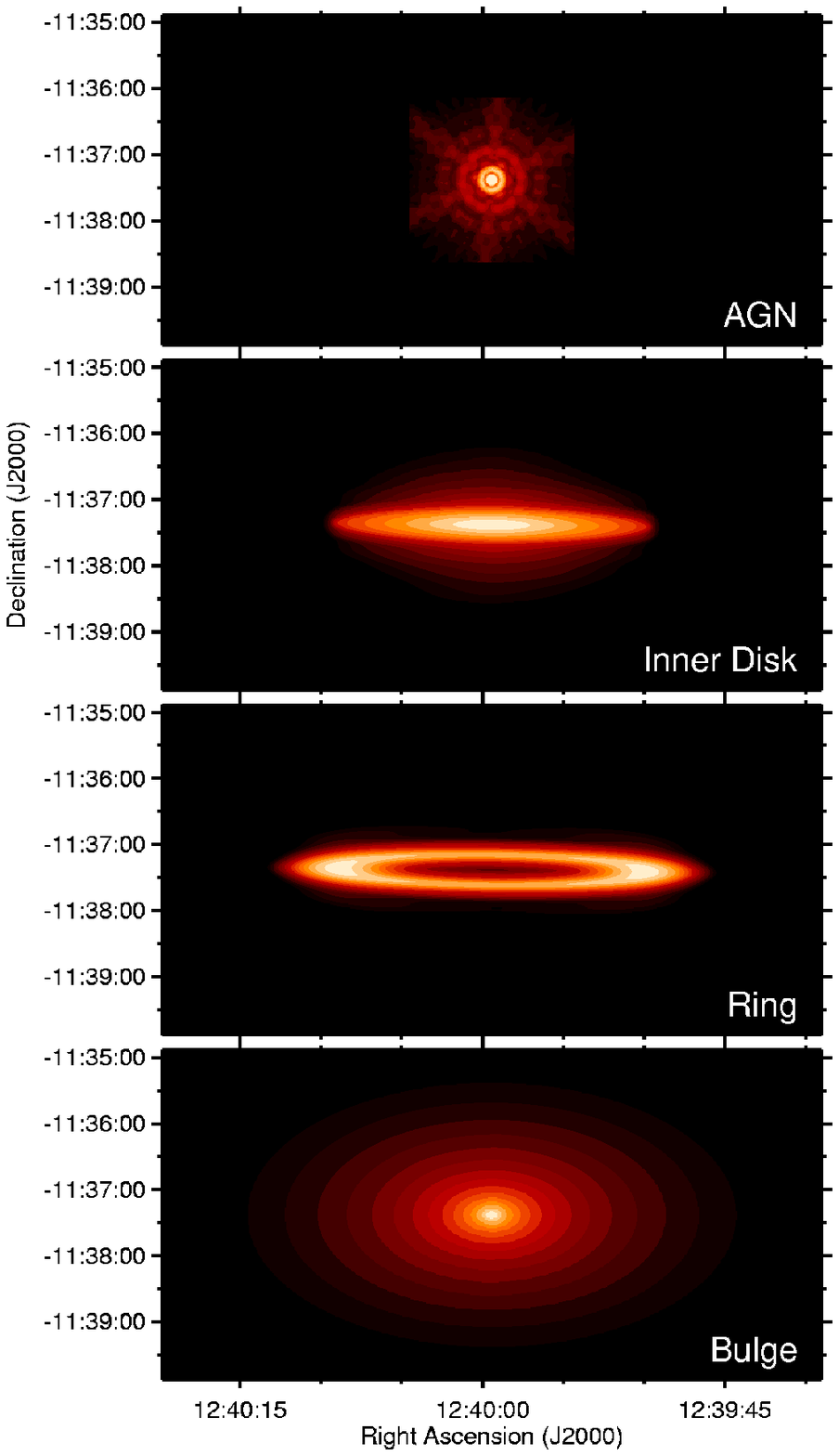}}
\caption{Models of the AGN, inner disk, ring, and bulge (all convolved
with the 24~$\mu$m model PSF) made using the parameters determined
from fitting the models to the 24~$\mu$m image.  This is presented as
an example of the image models that were fit to the data.  For display
purposes, the brightnesses of the model components are not scaled
relative to each other.}
\label{f_modelparts}
\end{figure}

\clearpage

\begin{figure}
\center{\includegraphics[height=8in]{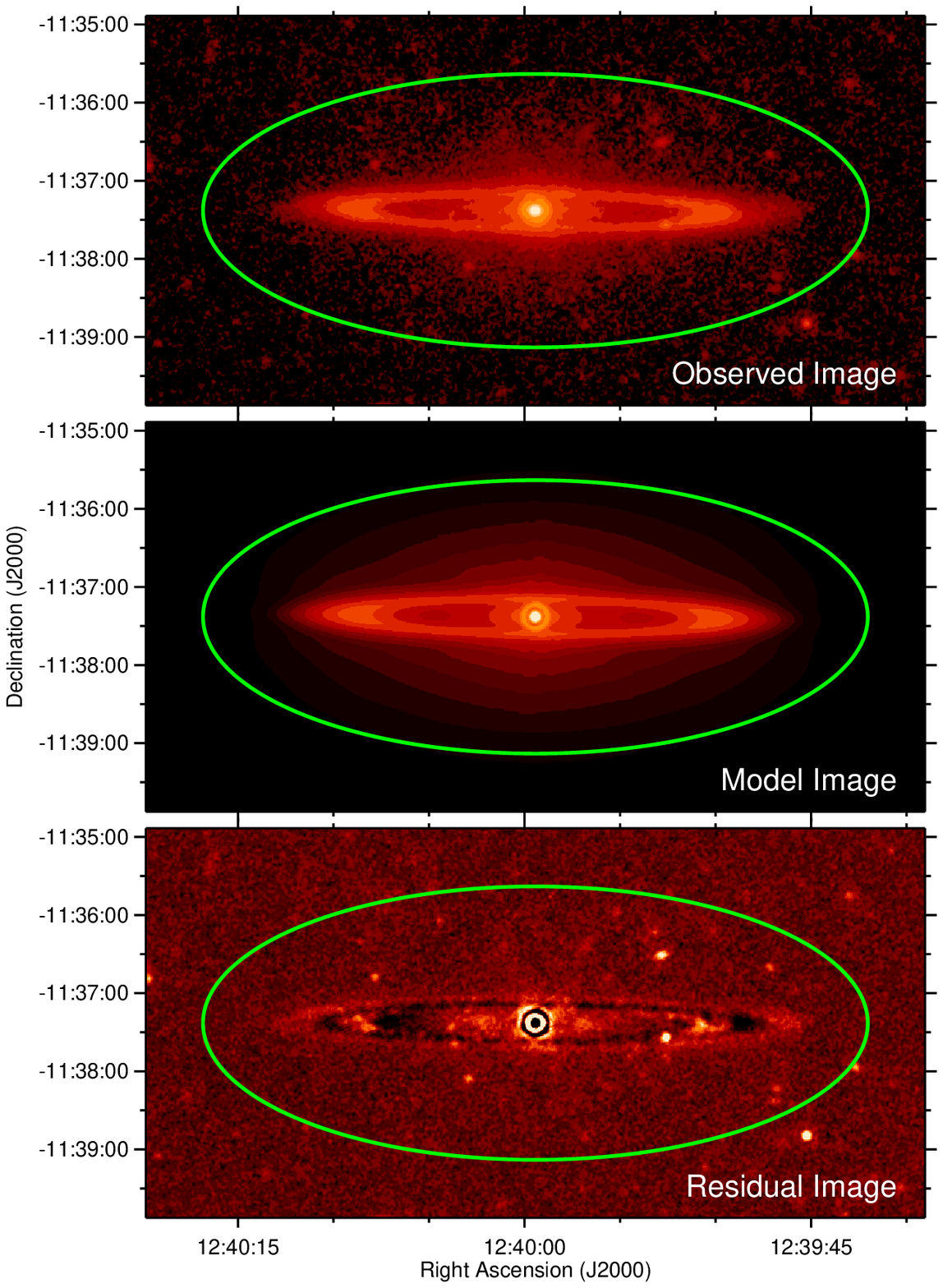}}
\caption{The observed 24~$\mu$m image, the model 24~$\mu$m image, and
the residuals from subtracting the model from the observed image.
This is presented as an example of the image models that were fit to
the data.  The optical disk of the galaxy (as given by
\citet{ddcbpf91}) is overplotted as green ovals to show the region
over which the fit was performed.}
\label{f_modelcompare}
\end{figure}

\clearpage

\begin{figure}
\center{\includegraphics[height=7in]{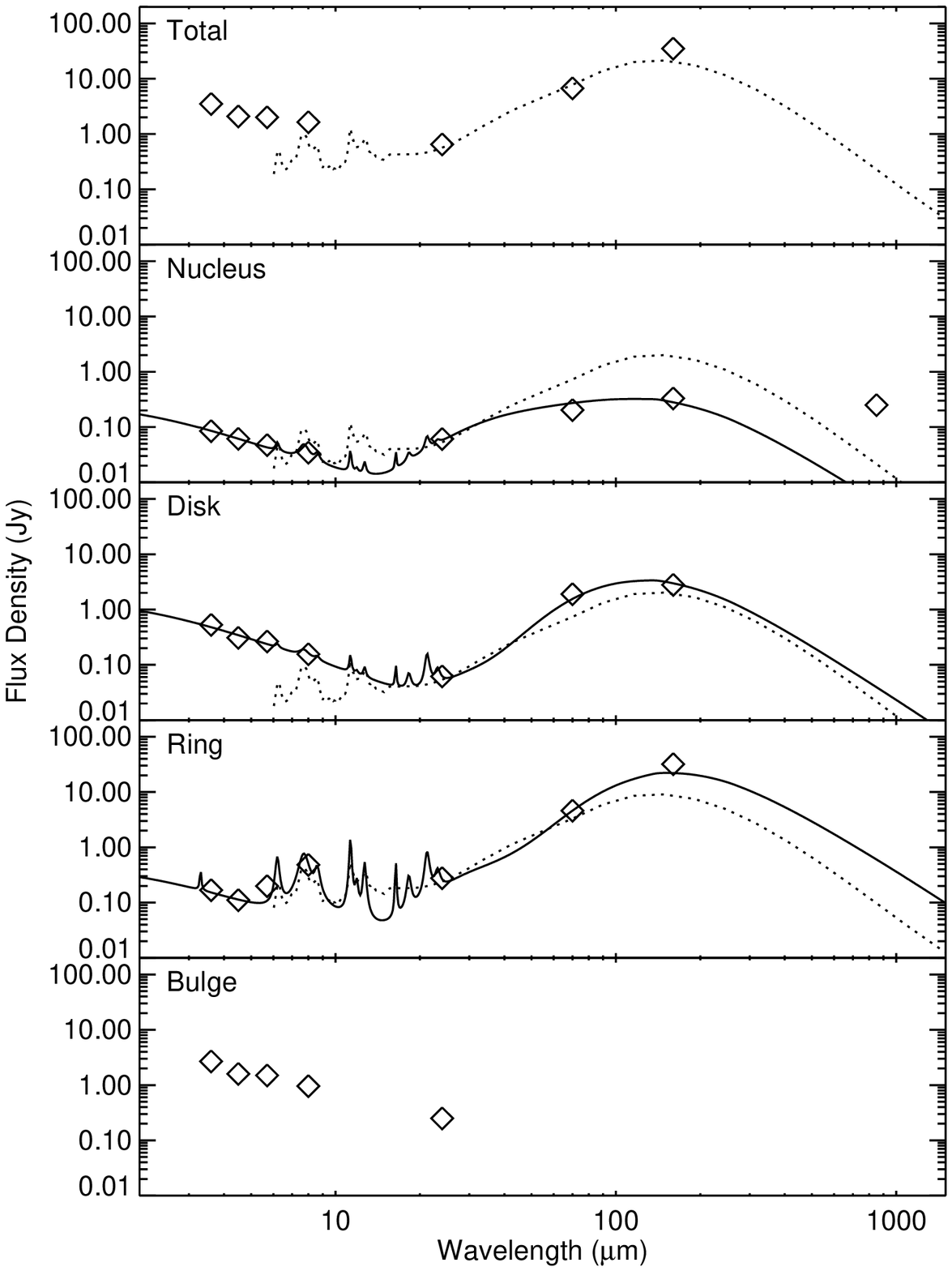}}
\caption{The SEDs of the separate components of NGC~4594.  The error
bars are equal to or smaller than the size of the plot symbols.  For
comparison between the SEDs, the y-axes of all the plots are fixed to
the same scale.  For comparison to the SEDs of nearby galaxies, the
semi-empirical dust model from \citet{detal05} with a median $\alpha$
value (explained in the text) is plotted over the total, nucleus,
disk, and ring SEDs as a dotted line.  This semi-empirical model has
been scaled to match the 24~$\mu$m flux density in each SED.  Emission
models comprised of starlight added to the \citet{ld01} dust models
are included in the plots of the nucleus, disk, and ring SEDs.}
\label{f_sed}
\end{figure}

\clearpage

\begin{figure}
\plotone{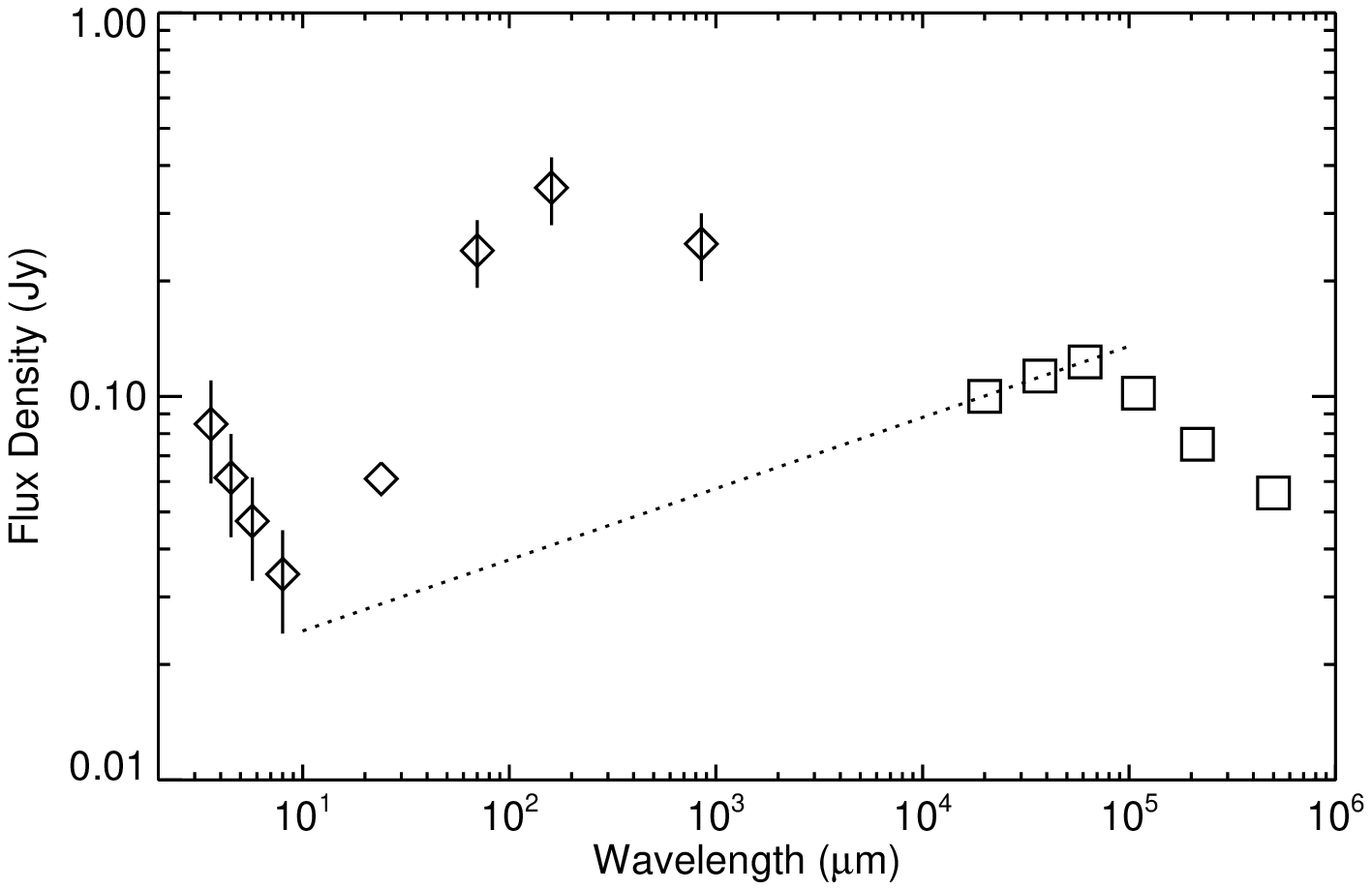}
\caption{The observed SED of the AGN of NGC~4594 (shown as diamonds),
along with the radio observations by \citet{hvd84} (shown as squares),
and the power law for the synchrotron emission as inferred from the
radio data (shown as a dotted line).  Except where shown, the error
bars are equal to or smaller than the size of the plot symbols.  Note
that the extrapolations fall short of the measured 850~$\mu$m flux
density.  See the text for a discussion of radio observations other
than those in \citet{hvd84}.}
\label{f_sed_agn}
\end{figure}

\clearpage

\begin{deluxetable}{lc}
\tablecolumns{2}
\tablewidth{0pc}
\tablecaption{Measurement of Select Spectral Lines in the Nucleus 
    of NGC~4594 \label{t_spectralmeas}}
\tablehead{
    \colhead{Feature} &
    \colhead{Measurement}}
\startdata
6.2~$\mu$m PAH feature equivalent width &     
    $0.0023 \pm 0.0004$~$\mu$m \\
6.2~$\mu$m PAH feature/continuum ratio &     
    $0.035 \pm 0.07$ \\
7.7~$\mu$m PAH feature equivalent width &     
    $< 0.006$~$\mu$m \\
7.7~$\mu$m PAH feature/continuum ratio &     
    $< 0.04$\\
12.8~$\mu$m [Ne~{\small II}] line strength &   
    $(1.86 \pm 0.10) \times 10^{-13}$ erg cm$^{-2}$ s$^{-1}$ \\
25.9~$\mu$m [O~{\small IV}] line strength &    
    $(5.21 \pm 0.17) \times 10^{-14}$ erg cm$^{-2}$ s$^{-1}$ \\
34.8~$\mu$m [Si~{\small II}] line strength &   
    $(7.2 \pm 0.2) \times 10^{-14}$ erg cm$^{-2}$ s$^{-1}$ \\
\enddata 
\end{deluxetable}

\clearpage

\begin{deluxetable}{cccccccc}
\tablecolumns{8}
\tablewidth{0pc}
\tabletypesize{\scriptsize}
\rotate
\tablecaption{Spatial Model Parameters of the Separate Components of NGC 4594 
    from Fits to Data\label{t_fit_spatial}}
\tabletypesize{\scriptsize}
\tablehead{ 
    \colhead{Wavelength} & 
    \colhead{Disk / Ring / Bulge} &
    \colhead{Disk / Ring} &
    \colhead{Inner Disk} &
    \multicolumn{2}{c}{Ring Parameters} &  
    \multicolumn{2}{c}{Bulge Scaling Parameter} \\
\colhead{} & 
    \colhead{Major Axis} &       \colhead{Minor/Major} &
    \colhead{Scale} &
    \colhead{Radius\tablenotemark{b}} &           
    \colhead{Width\tablenotemark{b}} &
    \colhead{Major} &            \colhead{Minor} \\
\colhead{} &
    \colhead{Position Angle\tablenotemark{a}} & 
    \colhead{Axis Ratio} &
    \colhead{Length\tablenotemark{b}} &
    \colhead{} &                 \colhead{} &
    \colhead{Axis} &             \colhead{Axis}
    }
\startdata

\hline
&
    &      &
    $50^{\prime\prime}.5 \pm 0^{\prime\prime}.5$ &
    &      &
    $141^{\prime\prime} \pm 4^{\prime\prime}$ &      
    $78.^{\prime\prime}.0 \pm 1^{\prime\prime}.1$ \\
\raisebox{1.5ex}[0pt]{3.6 $\mu$m} &
    \raisebox{1.5ex}[0pt]{$89^\circ.4 \pm 0^\circ$.4} &      
    \raisebox{1.5ex}[0pt]{$0.0996 \pm 0.0003$} &
    $2.25 \pm 0.02$ kpc&
    \raisebox{1.5ex}[0pt]{\tablenotemark{c}} &      
    \raisebox{1.5ex}[0pt]{\tablenotemark{c}} &
    $6.29 \pm 0.18$ kpc &      $3.48 \pm 0.05$ kpc \\

\hline
&
    &      &
    $51^{\prime\prime}.3 \pm 0^{\prime\prime}.9$ & 
    &      &
    $139^{\prime\prime} \pm 5^{\prime\prime}$ &
    $80^{\prime\prime} \pm 5^{\prime\prime}$ \\
\raisebox{1.5ex}[0pt]{4.5 $\mu$m} &
    \raisebox{1.5ex}[0pt]{$89^\circ.4 \pm 0^\circ.4$} &      
    \raisebox{1.5ex}[0pt]{$0.1010 \pm 0.0018$} &
    $2.29 \pm 0.04$ kpc &
    \raisebox{1.5ex}[0pt]{\tablenotemark{c}} &      
    \raisebox{1.5ex}[0pt]{\tablenotemark{c}} &
    $6.2 \pm 0.2$ kpc &   $3.6 \pm 0.2$ kpc \\

\hline
&
    &      &
    $54^{\prime\prime} \pm 3^{\prime\prime}$ &       
    $145^{\prime\prime}.9 \pm 1^{\prime\prime}$.6 &
    $24^{\prime\prime}.5 \pm 0^{\prime\prime}.3$ &
    $170^{\prime\prime} \pm 4^{\prime\prime}$ & 
    $102^{\prime\prime}.5 \pm 1^{\prime\prime}.6$ \\
\raisebox{1.5ex}[0pt]{5.7 $\mu$m} &
    \raisebox{1.5ex}[0pt]{$89^\circ.24 \pm 0^\circ.03$} &
    \raisebox{1.5ex}[0pt]{$0.1026 \pm 0.0010$} &
    $2.4 \pm 0.13$ kpc &       
    $6.51 \pm 0.07$ kpc &        $1.092 \pm 0.013$ kpc &
    $7.58 \pm 0.18$ kpc &        $4.57 \pm 0.07$ kpc \\

\hline
&
    &      &
    $46^{\prime\prime} \pm 3^{\prime\prime}$ & 
    $144^{\prime\prime} \pm 2^{\prime\prime}$ & 
    $21^{\prime\prime}.3 \pm 0^{\prime\prime}.8$ &
    $180^{\prime\prime} \pm 9^{\prime\prime}$ &   
    $101^{\prime\prime}.1 \pm 1^{\prime\prime}.2$ \\
\raisebox{1.5ex}[0pt]{8 $\mu$m} &
    \raisebox{1.5ex}[0pt]{$89^\circ.25 \pm 0^\circ.14$} &
    \raisebox{1.5ex}[0pt]{$0.099 \pm 0.003$} &
    $2.05 \pm 0.13$ kpc & 
    $6.42 \pm 0.09$ kpc &        $0.95 \pm 0.04$ kpc &
    $8.0 \pm 0.4$ kpc &          $4.5 \pm 0.05$ kpc \\

\hline
&
    &      &
    $45^{\prime\prime} \pm 7^{\prime\prime}$ & 
    $144^{\prime\prime}.9 \pm 1^{\prime\prime}.5$ &        
    $19^{\prime\prime}.4 \pm 0^{\prime\prime}.5$ &
    $180^{\prime\prime} \pm 50^{\prime\prime}$ &   
    $100^{\prime\prime} \pm 16^{\prime\prime}$ \\
\raisebox{1.5ex}[0pt]{24 $\mu$m} &
    \raisebox{1.5ex}[0pt]{$89^\circ.27 \pm 0^\circ.04$} &
    \raisebox{1.5ex}[0pt]{$0.0959 \pm 0.0009$} &
    $2.0 \pm 0.3$ kpc &       
    $6.46 \pm 0.07$ kpc &        $0.87 \pm 0.02$ kpc &
    $8 \pm 2$ kpc &              $4.5 \pm 0.7$ kpc \\

\hline
&
    &      &
    $39^{\prime\prime} \pm 5^{\prime\prime}$ &       
    $143^{\prime\prime}.9 \pm 1^{\prime\prime}.4$  &       
    $20^{\prime\prime}.1 \pm 1^{\prime\prime}.6$ &    
    &      \\
\raisebox{1.5ex}[0pt]{70 $\mu$m}  &
    \raisebox{1.5ex}[0pt]{$89^\circ.5 \pm 0^\circ.3$} &
    \raisebox{1.5ex}[0pt]{$0.102 \pm 0.005$} &  
    $1.7 \pm 0.2$ kpc &       
    $6.42 \pm 0.06$ kpc &       $0.90 \pm 0.07$ kpc &    
    \raisebox{1.5ex}[0pt]{\nodata} &
    \raisebox{1.5ex}[0pt]{\nodata} \\

\hline
\enddata 
\tablenotetext{a}{This angle is in degrees east from the North-South 
    axis.}
\tablenotetext{b}{Angular values for these paramteres refer to angular 
    distances along the major axis.}
\tablenotetext{c}{Values were fixed to weighted mean values from 
    5.7 - 70~$\mu$m data given in Table~\ref{t_fit_diskring}.}
\end{deluxetable}
%Note: 1''=44.6029 pc for dist=9.2 Mpc
%(Old system: 1''=66.4195 pc  for dist=13.7 Mpc)

\clearpage

\begin{deluxetable}{cc}
\tablecolumns{2}
\tablewidth{0pc}
\tablecaption{Weighted Mean and Standard Deviation of the Mean of Disk 
    and Ring Model Parameters from 5.7 - 70~$\mu$m data \label{t_fit_diskring}}
\tablehead{\colhead{Quantity} &                        \colhead{Value}}
\startdata

\hline
Disk/Ring Major Axis Position Angle\tablenotemark{a}&      
                               $89^\circ.25 \pm 0^\circ.06$ \\

\hline
Disk/Ring Major/Minor Axis Ratio &         
                               $0.0990 \pm 0.0015$ \\

\hline
&                              $48^{\prime\prime} \pm 3^{\prime\prime}$ \\
\raisebox{1.5ex}[0pt]{Inner Disk Scale Length\tablenotemark{b}}&
                               $2.14 \pm 0.13$ kpc \\

\hline
&                              $144^{\prime\prime}.7 \pm 0^{\prime\prime}.5$ \\
\raisebox{1.5ex}[0pt]{Ring Radius\tablenotemark{b}} &
                               $6.45 \pm 0.02$ kpc \\

\hline
&                              $22^{\prime\prime}.9 \pm 1^{\prime\prime}.1$ \\
\raisebox{1.5ex}[0pt]{Ring Width\tablenotemark{b}} &
                               $1.02 \pm 0.05$ kpc \\

\hline
\enddata 
\tablenotetext{a}{This angle is in degrees east from the North-South 
    axis.}
\tablenotetext{b}{Angular values for these paramteres refer to angular 
    distances along the major axis.}
\end{deluxetable}

\clearpage

\begin{deluxetable}{ccccccc}
\tablecolumns{7}
\tablewidth{0pc}
\tablecaption{Flux Density Model Parameters of the Separate Components of 
    NGC 4594  \label{t_fit_fd}}
\tabletypesize{\scriptsize}
\tablehead{ 
    \colhead{Wavelength} & 
    \multicolumn{5}{c}{Flux Density (Jy)\tablenotemark{a}} &  
    \colhead{Calibration} \\
    \colhead{($\mu$m)} & 
    \colhead{Global} &   \colhead{Nucleus} &   \colhead{Inner Disk} &
    \colhead{Ring} &     \colhead{Bulge} &
    \colhead{Uncertainty (\%)}
    }
\startdata
3.6 &   3.50 &
    $0.0847 \pm 0.0008$ &  $0.529 \pm 0.006$ &    
    $0.168 \pm 0.011$ &    $2.684 \pm 0.017$ &
    30\\
4.5 &   2.09 &
    $0.0614 \pm 0.0008$ &  $0.309 \pm 0.009$ &
    $0.111 \pm 0.003$ &    $1.595 \pm 0.016$ &
    30\\
5.7 &   2.01 &
    $0.0473 \pm 0.0003$ &  $0.265 \pm 0.006$ &
    $0.199 \pm 0.004$ &    $1.496 \pm 0.003$ &
    30\\
8 &     1.64 &
    $0.0344 \pm 0.0009$ &  $0.156 \pm 0.009$ &   
    $0.482 \pm 0.008$ &    $0.96 \pm 0.02$ &
    30\\
24 &    0.65 &
    $0.061 \pm 0.006$ &    $0.062 \pm 0.014$ &
    $0.277 \pm 0.006$ &    $0.25 \pm 0.03$ &
    10\\ 
70 &    6.7 &    
    $0.203 \pm 0.018$ &    $1.9 \pm 0.2$ &    
    $4.6 \pm 0.2$ &        \nodata &    
    20\\
160 &   35.1 &
    $0.33 \pm 0.05$ &      $2.8 \pm 0.2$ &
    $32.0 \pm 0.2$ &       \nodata &
    20\\
850 &   \nodata & 
    $0.25 \pm 0.05$ &      \nodata & 
    \nodata &              \nodata &
    10\\
\enddata 
\tablenotetext{a}{The uncertainties in these columns are from the fits
   to the data and do not include the calibration uncertainties.  The 
   calibration uncertainties are listed in the final column.}
\end{deluxetable}

\clearpage

\end{document}